\documentclass[a4paper,12pt]{article}
\usepackage[unicode]{hyperref}
\usepackage{subfigure}
\usepackage{amssymb, graphics}
\usepackage{pslatex}
\usepackage{ae}
\usepackage{mathrsfs}
\usepackage{graphicx}
\usepackage{bbm}
\usepackage{latexsym}
\usepackage{dsfont}
\usepackage{amsfonts}
\usepackage{amsmath}
\usepackage[title,titletoc]{appendix}
\usepackage[super]{nth}
\usepackage{cancel}

\usepackage{xcolor}
\usepackage{cite}
\usepackage{bbold}
\usepackage{booktabs}
\usepackage{stackengine}
\usepackage{mathtools}

\allowdisplaybreaks

\newcommand{\mathsym}[1]{{}}

\newcommand{\qed}{\nobreak \ifvmode \relax \else \ifdim\lastskip<1.5em \hskip-\lastskip \hskip1.5em plus0em minus0.5em \fi \nobreak \vrule height0.75em width0.5em depth0.25em\fi}
\newcommand{\diag}{\mbox{diag}}
\newcommand{\tr}{\mbox{tr}}

\def\app#1#2{  \mathrel{    \setbox0=\hbox{$#1\sim$}    \setbox2=\hbox{      \rlap{\hbox{$#1\propto$}}      \lower1.1\ht0\box0    }    \raise0.25\ht2\box2  }}

\DeclareMathOperator{\im}{Im}
\DeclareMathOperator{\re}{Re}

\numberwithin{equation}{section}

\newcommand{\id}{\mathds{1}}
\newcommand\brabar{\scalebox{.3}{(}\raisebox{-2pt}{\bf --}\scalebox{.3}{)}}

\begin{document}
\begin{titlepage} 
${}$
\begin{center} 
\vskip 1cm

{\large \bf
Type-I Seesaw with eV-Scale Neutrinos
} \vskip 1cm 

G.~C.~Branco  $^{a,b}$ \footnote{E-mail: \texttt{gbranco@tecnico.ulisboa.pt}},
J.~T.~Penedo  $^{a}$ \footnote{E-mail: \texttt{joao.t.n.penedo@tecnico.ulisboa.pt}},
Pedro~M.~F.~Pereira  $^{a}$ \footnote{E-mail: \texttt{pedromanuelpereira@tecnico.ulisboa.pt}},\\[1mm]
M.~N.~Rebelo  $^{a,b}$ \footnote{E-mail: \texttt{rebelo@tecnico.ulisboa.pt}. 
On leave of absence from $^{a}$ CFTP/IST, U. Lisboa},
and
J.~I.~Silva-Marcos  $^{a}$ \footnote{E-mail: \texttt{juca@cftp.tecnico.ulisboa.pt}}\\
\vskip 0.07in 

$^a$ Centro de
F{\'\i}sica Te\'orica de Part{\'\i}culas, CFTP,  Departamento de
F\'{\i}sica,\\ {\it Instituto Superior T\'ecnico, Universidade de Lisboa, }
\\ {\it Avenida Rovisco Pais nr. 1, 1049-001 Lisboa, Portugal;} \\

$^b$ CERN, Theoretical Physics Department, 
{\it CH-1211 Geneva 23, Switzerland.}

\end{center}

\vskip 1cm
\begin{abstract} 

We consider seesaw type-I models including at least one (mostly-)sterile neutrino with mass at the eV scale. Three distinct situations are found, where the presence of light extra neutrinos is naturally justified by an approximately conserved lepton number symmetry. To analyse these scenarios consistently, it is crucial to employ an exact parametrisation of the full mixing matrix. We provide additional exact results, including generalised versions of the seesaw relation and of the Casas-Ibarra parametrisation, valid for every scale of seesaw. We find that the existence of a light sterile neutrino imposes an upper bound on the lightest neutrino mass. We further assess the impact of light sterile states on short- and long-baseline neutrino oscillation experiments, emphasise future detection prospects, and address CP Violation in this framework via the analysis of CP asymmetries and construction of weak basis invariants. The proposed models can accommodate enough active-sterile mixing to play a role in the explanation of short-baseline anomalies.

\end{abstract}
\end{titlepage}
\setcounter{footnote}{0}

\newpage

\section{Introduction}

Most of the present data on Neutrino Physics are consistent with the
hypothesis of having only three active neutrinos. Nevertheless, there is a
small subset of experiments which seem to require the presence of New
Physics (NP). The first indication hinting at the presence of NP was
provided by an excess in the results of the LSND experiment,
where electron anti-neutrinos were observed in a
pure muon anti-neutrino beam \cite{Athanassopoulos:1996jb,Aguilar:2001ty}.
One of the simplest explanations of the LSND result
involves the existence of an anti-neutrino with a mass-squared difference $\Delta m^{2}$ of about 1
eV$^{2}$. Taking into account that $\Delta m^{2}_\text{atm}$ is of order 
$10^{-3}$ eV$^{2}$ and $\Delta m^{2}_\text{solar}$ of order \mbox{$10^{-4}$ eV$^{2}$}
one concludes that the LSND result would require a fourth neutrino. On
the other hand, the invisible decay of the $Z$ gauge boson shows that there
are only three active neutrinos with a mass less than a half of the $Z$
mass~\cite{PDG2019}, implying that if a fourth light neutrino exists it must be sterile,
i.e., a singlet under the gauge symmetry of the Standard Model (SM).
The existence of extra (sterile) neutrinos 
should then be reconciled with cosmological constraints, which call for a suppressed
thermalisation of these massive neutrinos in the early Universe,
given the effective neutrino number $N_\text{eff} = 2.99^{+0.34}_{-0.33}$
($95\%$~CL, from TT,TE,EE+lowE+lensing+BAO) measured by Planck~\cite{Aghanim:2018eyx}%
\footnote{
Although addressing this suppression falls beyond our scope, it has
been shown that it can be achieved via ``secret'' sterile neutrino
self-interactions~\cite{Hannestad:2013ana,Dasgupta:2013zpn}. Here, TT,TE and EE+lowE+lensing refer to particular likelihood combinations and BAO stands for baryon acoustic oscillation measurements.
}.

Meanwhile, new anomalies have appeared in Neutrino Physics supporting the
hypothesis of the existence of light sterile neutrinos. The indications for
the existence of a sterile neutrino of mass of order 1 eV come from
short-baseline (SBL) neutrino oscillation experiments.
They started with the LSND
result in the nineties. At that time, this result was not confirmed by
KARMEN \cite{Armbruster:2002mp}. However, KARMEN had a shorter baseline than
LSND and therefore could not exclude the whole parameter space available to
LSND. This was followed by the MiniBooNE experiment \cite{Aguilar-Arevalo:2018wea}
with inconclusive results%
\footnote{
The need to reconcile MiniBooNE and LSND data has currently revived
interest~\cite{Dentler:2019dhz,deGouvea:2019qre} in models attempting to explain
anomalies via sterile neutrino decay~\cite{PalomaresRuiz:2005vf}.}. Recently, new interest
in the LSND result was sparked by the ``reactor anti-neutrino anomaly'' due to
a deficit of the number of anti-neutrinos observed in several different
reactor neutrino experiments, when compared with the theoretical flux
calculations \cite{Mueller:2011nm,Mention:2011rk,Huber:2011wv}.
A crucial and independent development has been provided by the DANSS~\cite{Alekseev:2018efk}
and NEOS~\cite{Ko:2016owz} collaborations,
whose programmes include comparing spectra at different distances from the anti-neutrino source.
The preferred fit regions of these independent experiments interestingly overlap near $\Delta m^2 \sim 1.4$ eV$^2$
and $\sin^2 2 \vartheta_{14} \sim 0.05$, with $\vartheta_{14}$ being an effective mixing angle as interpreted in a 3+1 scheme.
Also of relevance is the so-called ``Gallium neutrino anomaly'',
discovered in 2005-2006~\cite{Abdurashitov:2005tb,Laveder:2007zz,Giunti:2006bj},
albeit of less significance.
For recent reviews on eV-scale sterile neutrinos and additional references, see~\cite{Giunti:2019aiy,Diaz:2019fwt}. 

The purpose of this paper is to investigate the possibility of obtaining in a natural way
at least one sterile neutrino with a mass of order eV in the framework of the
general type-I seesaw mechanism~\cite{Minkowski:1977sc,Yanagida:1979as,Glashow:1979nm,GellMann:1980vs,Mohapatra:1979ia}.
The crucial point is that we
shall consider a special case of the seesaw framework. Instead of having
three heavy sterile neutrinos, as in the usual setup,
at least one of the sterile neutrinos should be light while, at the same time, its
mixing with the light active neutrinos should be small enough to comply with existing
experimental bounds, but large enough to be relevant to low energy
phenomenology. Two important challenges are to find solutions that are
stable under renormalisation, and to inquire if these spectra,
with at least one neutrino with a mass of order eV,
might indeed explain the SBL anomalies.

For definiteness, let us recall how the conventional seesaw mechanism
works. It consists of an extension of the SM where three right-handed
neutrinos are added to the standard spectrum. As a result, the neutrino mass
terms include a Dirac mass matrix, denoted $m$, generated by the breakdown
of the electroweak (EW) symmetry, and a Majorana mass term, denoted $M$, with the
scale of $M$ much larger than the scale of $m$. In general this leads to
three light neutrinos with masses of order $m^{2}/M$ and three heavy neutrinos
with masses of order $M$. The generic seesaw framework leads to an active-sterile mixing 
of order $m/M$, too small to be of relevance to low
energy physics, while providing a framework for leptogenesis~\cite{Fukugita:1986hr}.
In the derivation of the standard seesaw formulae, one
performs a block diagonalisation of the $6 \times 6$ complex neutrino mass matrix,
obtaining approximate relations that are valid to an excellent
approximation. Some of the approximate formulae no longer hold in the
special cases which we are considering. However, there are important exact
relations which continue to be valid in our case. We find viable models with
at least one sterile neutrino with a mass of order eV by imposing a $U(1)$
symmetry (see e.g.~\cite{Branco:1988ex}) allowing for small breaking terms.
Before the breaking, for special assignments of leptonic charges, the lightest neutrinos are
naturally massless at tree level, acquiring calculable small masses after
the breaking and complying with the experimental $\Delta m^2$ values after
radiative corrections.

The paper is organised as follows. In the next section, we describe our setup,
settle the notation and present a useful parametrisation of the mixing matrix
as well as some exact results concerning the Dirac mass matrix, neutrino masses and deviations from unitarity.
In section~\ref{sec:devunit} we discuss the size of such deviations
from unitarity in the $3\times 3$ leptonic mixing matrix.
In section~\ref{sec:loopandsym} we describe how one-loop mass corrections can be controlled
within the considered framework. In section~\ref{sec:numeric} we present explicit numeric examples
and go through their phenomenology,
while section~\ref{sec:cp} is dedicated to the study of CP Violation within the type-I seesaw,
with emphasis on CP Violation measurements and CP-odd weak basis invariants.
Finally our conclusions are presented in section~\ref{sec:conclusions}.

\section{Framework}
\label{sec:framework}

We work under the type-I seesaw framework, in a model with three
right-handed neutrinos added to the SM.
The leptonic mass terms are given by: 
\begin{equation}
\begin{aligned}
\mathcal{L}_{m} \,&=\,
-\left[\overline{{\nu }_{L}^{0}}\,m\,\nu _{R}^{0}
+\frac{1}{2}\nu_{R}^{0T} C^{*} M\,\nu _{R}^{0}
+\overline{l_{L}^{0}}\,m_{l}\,l_{R}^{0}\right]+\text{h.c.} \\
\,&=\,
-\left[\frac{1}{2}n_{L}^{0T} C^{*} \mathcal{M}^* n_{L}^{0}
+\overline{l_{L}^{0}} m_{l}l_{R}^{0}\right]+\text{h.c.}\,, 
\end{aligned}
\end{equation}
where $n_L^0 = (\nu_L^0\,, \,\,C\, \overline{\nu_R^0}^T)^T$
and the zero superscript denotes a general flavour basis.
Without loss of generality, one may choose a weak basis where $m_{l}$ is
real and diagonal. The analysis that follows is performed in this
basis, meaning $\nu_L^0 = (\nu_{eL},\,\nu_{\mu L},\,\nu_{\tau L})$.
The neutrino mass matrix $\mathcal{M}$ is a $6\times 6$ complex symmetric matrix and
has the form: 
\begin{align}
\mathcal{M}=\left( 
\begin{array}{cc}
0 & m \\ 
m^{T} & M%
\end{array}%
\right)\,.  \label{vm0}
\end{align}%
This matrix is diagonalised by the unitary transformation 
\begin{align}
\mathcal{V}^{T}\mathcal{M}^*\mathcal{V}=\mathcal{D}\qquad
\Longleftrightarrow \qquad \mathcal{M}=\mathcal{V\ D\ V}^{T},  \label{vm}
\end{align}%
where $\mathcal{D}$ is diagonal real non-negative
and contains all neutrino masses,
\begin{align}
\mathcal{D}=\left( 
\begin{array}{cc}
d & 0 \\ 
0 & D%
\end{array}%
\right)\,.  \label{eq:d}
\end{align}%
Here, $d$ contains the masses of the three known light neutrinos, $d=\text{diag}(m_1,m_2,m_3)$,
and $D$ the masses of other
neutrinos, $D=\text{diag}(M_1,M_2,M_3)$. The $6\times 6$
unitary matrix $\mathcal{V}$ can be written as
\begin{align}
\mathcal{V}=\left( 
\begin{array}{cc}
K & R \\ 
S & Z%
\end{array}%
\right) \,,   \label{unit}
\end{align}%
where $K$, $R$, $S$ and $Z$ are $3\times 3$ matrices. Using
the unitarity of $\mathcal{V}$, namely $\mathcal{V}\,\mathcal{V}^\dagger=\mathcal{V}^\dagger\mathcal{V}=\id_{(6\times 6)}$,
one can obtain~\cite{Agostinho:2017wfs} a series of exact relations relating
the matrices $K$, $R$, $S$, and $Z$,
examples of which are $KK^{\dagger }+RR^{\dagger }=\id$ and $KS^{\dagger }+RZ^{\dagger }=0$.
We shall show that in order to study deviations of unitarity,
it is useful to parametrise $\mathcal{V}$ in a different way.

\subsection{A Novel Parametrisation for the Leptonic Mixing Matrix}
\label{sec:param}
In Ref.~\cite{Agostinho:2017wfs} we introduced an especially useful parametrisation of the $6 \times 6$
leptonic mixing matrix that enables to control all deviations from unitarity through a single $3 \times 3$
matrix which connects the mixing of the active and sterile neutrinos in the context of type I seesaw.
It was written:
\begin{align}
\mathcal{V}=\left( 
\begin{array}{cc}
K & 0 \\ 
0 & Z%
\end{array}%
\right) \left( 
\begin{array}{cc}
\id\, & Y \\ 
-X & \id\,%
\end{array}%
\right) \,,\quad X=-Z^{-1}S\,,\quad Y=K^{-1}R\,,
\label{u2}
\end{align}
where it is assumed that $K$ and $Z$ are non-singular.
From the aforementioned unitarity relation $KS^{\dagger }+RZ^{\dagger }=0$
one promptly concludes that 
\begin{align}
Y=X^{\dagger }\quad \Longrightarrow \quad \mathcal{V}=\left( 
\begin{array}{cc}
K & KX^{\dagger } \\ 
-ZX & Z%
\end{array}%
\right)\,.  \label{xy}
\end{align}
Thus, a generic $6\times 6$\ unitary matrix $\mathcal{V}$, in fact, only
contains three effective $3\times 3$ matrices $K$, $Z$ and $X$. Furthermore,
from the same unitarity of $\mathcal{V}$
and from the singular value decomposition $X = W\,d_X\,U^\dagger$,
one finds that $K$ and $Z$ can be written as:
\begin{align}
\begin{array}{l}
K = U_{K}\ \sqrt{\left( \id\,+d_{X}^{2}\right) ^{-1}}\ U^{\dagger }
  = U_{K}\ U^{\dagger }\sqrt{\left( \id\,+X^{\dagger}X\right) ^{-1}}
= V \sqrt{\left( \id\,+X^{\dagger}X\right) ^{-1}}
\,, \\[4mm] 
Z = W_{Z}\ \sqrt{\left( \id\,+d_{X}^{2}\right)^{-1}}\ W^{\dagger }
  = W_{Z}\ W^{\dagger }\ \sqrt{\left( \id\,+XX^{\dagger}\right) ^{-1}}
\,,%
\end{array}
\label{kz}
\end{align}%
where%
\footnote{Principal square roots of positive semi-definite matrices
are unique and their use is implied in Eq.~\eqref{kz}.}
 $U_{K},W_{Z}$, $U$ and $W$ are all $3\times 3$ unitary matrices,
$d_X$ is a diagonal matrix with real non\discretionary{-}{-}{-}negative entries,
and we have defined an additional unitary matrix $V \equiv U_K U^\dagger$.
The matrices $U$ and $W$ diagonalise the Hermitian products
$X^{\dagger }X$ and $XX^{\dagger }$, respectively: 
\begin{align}
U^{\dagger }\ X^{\dagger }X\ U=d_{X}^{2}\,,\qquad W^{\dagger
}XX^{\dagger }\ W=d_{X}^{2} \,.
\label{uwxx}
\end{align}
Any unitary matrix to the left of $Z$ --
like the product $W_Z\,W^\dagger$ in Eq.~\eqref{kz} --
is unphysical as it can be rotated away via a weak
basis transformation which does not affect the form of $m_l$.
Accordingly, one can choose to work in a weak basis for which $\Sigma = \id$
in the general expression 
\begin{align}
Z\,=\,\Sigma \,( \id\,+XX^{\dagger})^{-1/2}\,,
\label{eq:Zgeneral}
\end{align}
with $\Sigma$ unitary. 
Note, however, that $\Sigma \neq \id$ in the numerical `symmetry' bases considered later on
in sections~\ref{sec:loopandsym} and~\ref{sec:numeric}.

The matrix $K$ plays the role of the PMNS mixing matrix,
as it connects the flavour eigenstates $\nu_{\alpha L}$
($\alpha = e$, $\mu$, $\tau$)
to the lightest mass eigenstates.
From Eq.~\eqref{kz}, it is clear that $K$ is unitary if and only if 
$d_{X}^{2}=0$. Thus, the deviations from unitarity are manifestly expressed
in the diagonal matrix $d_{X}^{2}$ containing the (squared) 
singular values of $X$.

In summary, a generic $6\times 6$ mixing unitary matrix $\mathcal{V}$ can
be simplified and
be written in terms of just one $3\times 3$ unitary matrix $V$
and of explicit deviations from unitarity, parametrised by a $3 \times 3$ matrix $X$:%
\begin{align}
\mathcal{V}=\left( 
\begin{array}{cc}
K & R \\ 
S & Z%
\end{array}%
\right) \qquad ;\qquad 
\begin{array}{l}
K=V\,
\sqrt{\left( \id +X^{\dagger
}X\right) ^{-1}}\qquad ;\qquad R=K\ X^{\dagger }, \\[5mm] 
Z=
\sqrt{\left( \id +XX^{\dagger
}\right) ^{-1}}\qquad \,\;\;\;;\qquad S=-Z\ X\,,
\end{array}
\label{eq:epl}
\end{align}%
i.e.
\begin{align}
\mathcal{V}=\left( 
\begin{array}{cc}
V\,\left( \id+X^\dagger X\right)^{-1/2} & 
V\,\left( \id+X^\dagger X\right)^{-1/2}\,X^\dagger \\ 
-\left( \id+XX^\dagger\right)^{-1/2}\, X & 
\left( \id+XX^\dagger\right)^{-1/2}
\end{array}%
\right)\,.
\end{align}

In general, there are no restrictions on the matrix $X$. However, in a type-I
seesaw model, the mixing matrix $\mathcal{V}$ must also obey the mass
relation stated in Eq.~\eqref{vm}, and the $6\times 6$ neutrino mass matrix $%
\mathcal{M}$ is not general: some entries are zero at tree level. This imposes
a restriction%
\footnote{This restriction generalises to $d + X^\dagger D X^*= K^{-1} m_L (K^{-1})^\dagger$
for an explicit, symmetric light neutrino Majorana mass matrix $m_L$ in place of the zero in Eq.~\eqref{vm0},
which may arise from radiative corrections or
be present due to e.g.~a type-II seesaw~\cite{Konetschny:1977bn, Schechter:1980gr, Cheng:1980qt,
Lazarides:1980nt, Mohapatra:1980yp} contribution.}
 on $X$,
\begin{align}
d+X^{T}\ D\ X=0\,,
\label{dXDX}
\end{align}%
which implies that it is possible to write $X$ as: 
\begin{align}
X\,=\,i\,\sqrt{D^{-1}}\,O_{c}\,\sqrt{d}\,,
\label{eq:xc}
\end{align}%
where $O_c$ is a complex orthogonal matrix, i.e., $O_c^T O_c=O_c\,O_c^T=\id$. 
Explicitly, 
\begin{align}
|X_{ij}|=\left\vert (O_c)_{ij}\,\sqrt{\frac{m_j}{M_i}}\,\right\vert\,.
\end{align}%

Since $O_{c}$ is an orthogonal complex matrix, not all of its elements need
to be small. Furthermore, not all the $M_{i}$ need to be much larger than
the electroweak scale, in order for the seesaw mechanism to lead to
naturally suppressed neutrino masses. These observations about the size of
the elements of $X$ are especially relevant in view of the fact that some of
the important physical implications of the seesaw model depend crucially on $X$.
In particular, the deviations of $3\times 3$ unitarity are controlled by 
$X$, as shown in Eq.~\eqref{kz}.
On the other hand, from Eq.~\eqref{dXDX} one can also see that $X$ must
not vanish, in order to account for the non-zero light neutrino 
masses. 
Several authors have adopted different types of parametrisations for the full mixing matrix, in the context of seesaw models, see for example \cite{Korner:1992zk,Casas:2006hf,Blennow:2011vn,Xing:2011ur,Donini:2012tt}.
Some of these are approximate and apply to specific limits or to models with fewer than three sterile neutrinos, others are exact and do not depend on the number of sterile neutrinos like in our case.%
\footnote{
Although we have applied our parametrisation to a scenario with three sterile neutrinos, it is applicable to cases where the number $q$ of sterile neutrinos differs from 3. We are then in the presence of a rectangular $3 \times q$ Dirac mass matrix $m$ and of a $q \times 3$ rectangular $X$ matrix, with everything else remaining consistent.
} 
Some of these parametrisations were derived to deal with special types of analyses and may become cumbersome when adopted for other purposes. We find our parametrisation very useful since it is particularly simple and parametrises, in a concise and exact form, all deviations from unitarity by a single matrix $X$.

From the above, one concludes that the set $\{m_l,d,D, V, O_c\}$
of matrices is sufficient to describe lepton masses and mixing
at tree level.
In the working weak basis,
there are 9 lepton masses in the first three matrices, while
mixing is parametrised by 6 angles and 6 CP-violating (CPV) phases,
contained in the unitary matrix $V$ and in the orthogonal deviation matrix $O_c$.~Parameter counting is summarised in Table~\ref{tab:parameters}
and is in agreement with, e.g., Refs.~\cite{Branco:2001pq, Broncano:2002rw, Branco:2011zb}.
Coincidentally, these numbers of angles and CPV phases match
those of a general 3+1 scenario (see e.g.~\cite{Giunti:2007ry}),
even though three right-handed neutrinos have been added to the SM.
This is a consequence of having a type-I seesaw UV completion,
which requires the zero block in Eq.~\eqref{vm0}.
\begin{table}[t]
\centering
\renewcommand{\arraystretch}{1.2}
\begin{tabular}{lcccccc}
\toprule 
 & $\,\,\,\, m_l \,\,\,\,$ & $\,\,\,\, d \,\,\,\,$ & $\,\,\,\, D \,\,\,\,$ & $\,\,\,\, O_c \,\,\,\,$ & $\,\,\,\, V \,\,\,\,$ & Total \\
\midrule
Moduli \quad& 3 & 3 & 3 & 3 & 3 & 15 = 9 + 6 \\[1mm] 
Phases \quad& $-3$ & 0 & 0 & 3 & 6 & 6 \\ 
\bottomrule
\end{tabular}
\caption{
Physical parameter counting in type-I seesaw with three sterile neutrinos. The 15 moduli correspond to 9 lepton masses
(3 charged-lepton masses and 6 neutrino masses) and to 6 mixing angles. There are 6 physical phases,
as rephasing the charged leptons can remove 3 phases from $V$.
Recall that $m_l$ is real and diagonal in the considered weak basis.}
\label{tab:parameters}
\end{table}

In this paper, we consider the possibility of having at least one sterile
neutrino with a mass of order eV arising from the seesaw mechanism in a
model with three right-handed neutrinos added to the SM.
We analyse the different aspects and consequences of the phenomenology of such a model. 
With this aim, relations between observables and parameters
which are independent of the seesaw limit are derived in the following subsection.

\subsection{Exact Relations at Tree Level}

From Eqs.~\eqref{vm} and \eqref{xy}, one can extract a general 
and exact formula for the neutrino Dirac mass matrix $m$ in Eq.~\eqref{vm0},
valid for any weak basis and any scale of $M$:
\begin{align}
m \,=\, K\, X^\dagger D \left( Z^{-1}\right)^* 
  \,=\, - i \, K \, \sqrt{d} \, O_{c}^\dagger \sqrt{D}
\left( Z^{-1}\right)^*.
\label{eq:mdirac}
\end{align}
Recall that, in our working weak basis,
$m_l$ is diagonal and
$K$ is directly identified with the 
non-unitary PMNS matrix.
Moreover, $K$ and $Z$ take the forms given in Eq.~\eqref{eq:epl}
and one has:
\begin{equation}
\begin{aligned}
m \,&=\,
V\,\sqrt{\left(\id+X^\dagger X\right)^{-1}}\,X^\dagger D\, \sqrt{\id+X^* X^T} \\
&=\, - i \,
V\,\sqrt{\left(\id+X^\dagger X\right)^{-1}}
\, \sqrt{d} \, O_{c}^\dagger \sqrt{D}\,
\sqrt{\id+X^* X^T}\,.
\end{aligned}
\label{eq:mdiracWB}
\end{equation}
This exact formula is to be contrasted with the known
parametrisation for the neutrino Dirac mass matrix developed by
Casas and Ibarra~\cite{Casas:2001sr}, which is
valid in the standard seesaw limit of $M \gg m$ and reads
\begin{align}
m\, \simeq \,-i\, U_\text{PMNS} \, \sqrt{d} \, O_c^\text{CI} \, \sqrt{D}\,,
\label{eq:casas}
\end{align}
in the weak basis where $m_l$ and $M = \diag(\tilde M_1, \tilde M_2, \tilde M_3) \equiv \tilde D$ are diagonal.
Here, $O_c^\text{CI}$ is an orthogonal complex matrix
and $U_\text{PMNS}$ represents the approximately unitary lepton mixing matrix.
In this limit of $M \gg m$, the light neutrino mass matrix $m_\nu$ can be approximated by:
\begin{align}
m_\nu\,\simeq \, - m\, M^{-1}m^T.
\label{eq:eff}
\end{align}
It is clear from \eqref{eq:mdiracWB} that
one can obtain Eq.~\eqref{eq:casas} as a limiting
case of Eq.~\eqref{eq:mdirac} 
through an expansion in powers of $X$.
Keeping only the leading term, unitarity is regained with $U_\text{PMNS} \simeq V$
and one can identify the complex orthogonal matrices: $O_c^\text{CI} = O_c^\dagger$.

As a side note, let us remark that it is possible to 
obtain a parametrisation for $m$ 
which is exact and holds in a general weak basis
by following the Casas-Ibarra procedure. One finds:
\begin{align}
m\, = \,-i\, U_\nu \, \sqrt{\tilde d} \, \tilde O_c^\text{CI} \, \sqrt{\tilde D}\,\, \Sigma_M^T,
\label{eq:casas_exact}
\end{align}
where once again $\tilde O_c^\text{CI}$ is a complex symmetric matrix. 
However, $\tilde d$ and $\tilde D$ do not contain physical masses,
but are instead
diagonal matrices with non-negative entries
obtained from the Takagi decompositions
$-m \,M^{-1} m \,=\, U_\nu \,\tilde d \,U_\nu^T$ and $M \,=\, \Sigma_M \,\tilde D \, \Sigma_M^T$,
with $U_\nu$ and $\Sigma_M$ unitary.
The matrix $\Sigma_M$ is unphysical, as it can be rotated away by a weak basis transformation
diagonalising $M$.
Even though this parametrisation resembles that of Eq.~\eqref{eq:mdiracWB},
the latter may be preferable since it directly makes use of low-energy observables.
Only in the limit $M\gg m$, where Eq.~\eqref{eq:eff} and $\tilde d \simeq d$, $\tilde D \simeq D$ hold,
does Eq.~\eqref{eq:casas_exact} reduce to the approximate relation \eqref{eq:casas},
in a weak basis of diagonal charged leptons and diagonal sterile neutrinos.

At this stage, one may wonder whether there exists an exact relation, analogous
to Eq.~\eqref{eq:eff} which is valid in any region of parameter space.
One can actually deduce such a relation for an arbitrary number of active and sterile neutrinos.
Consider the following decomposition of a block-diagonal matrix:
\begin{align}
\left[ 
\begin{array}{cc}
\mathbf{A} & \mathbf{B} \\ 
\mathbf{C} & \mathbf{D}
\end{array}\right]
\,=\,
\left[ 
\begin{array}{cc}
\id_{(p\times p)} & \mathbf{B} \\ 
0 & \mathbf{D}
\end{array}\right]
\left[ 
\begin{array}{cc}
\mathbf{A}-\mathbf{B}\,\mathbf{D}^{-1}\mathbf{C} & 0 \\ 
\mathbf{D}^{-1}\mathbf{C} & \id_{(q\times q)}
\end{array}\right]
\,,
\end{align}
where $\mathbf{A}$, $\mathbf{B}$, $\mathbf{C}$, and $\mathbf{D}$ are complex $p\times p$, $p\times q$, $q\times p$, and $q\times q$ matrices, respectively,
and one has assumed that $\mathbf{D}$ is non-singular.
From this it follows that
\begin{align}
\det \left[ 
\begin{array}{cc}
\mathbf{A} & \mathbf{B} \\ 
\mathbf{C} & \mathbf{D}
\end{array}%
\right] \,=\,
\det \left(\mathbf{A}-\mathbf{B}\,\mathbf{D}^{-1}\mathbf{C}\right) \,\, \det \,\mathbf{D}\,.
\label{po}
\end{align}%
In a general type-I seesaw scenario, $\mathbf{A}=0$, $\mathbf{B}=\mathbf{C}^T=m$ and $\mathbf{D}=M$, and one obtains%
\begin{align}
\left\vert \det \left[ 
\begin{array}{cc}
0 & m \\ 
m^{T} & M%
\end{array}%
\right] \right\vert =\left\vert \det\, m \right\vert ^{2}\,,
\label{po1}
\end{align}%
which leads to%
\begin{align}
m_{1}\ldots m_p=\frac{\left\vert \det \,m \right\vert ^{2}}{%
M_{1}\ldots M_q}\,,  \label{po21}
\end{align}%
with $m_{i}$ ($i=1,\ldots,p$) and $M_{j}$ ($j=1,\ldots,q$) denoting the neutrino masses.
For the case of interest, $p = q = 3$ and one has:
\begin{align}
m_1 \,m_2 \,m_3\,=\,\frac{\left\vert \det \,m \right\vert ^{2}}{%
M_1 \,M_2\, M_3}\,.  \label{po22}
\end{align}%
We stress that these relations are {\bf exact} and that no assumptions have been made about the
relative sizes of the $m_i$ and $M_j$.
It is clear from Eq.~\eqref{po22} that the smallness of neutrino masses in 
this framework may 
have its origin in the largeness of the $M_j$ (with respect to the EW scale),
or in the suppression of $|\det\, m|$ due to e.g.~an approximate symmetry.

\section{The Size of Deviations from Unitarity}
\label{sec:devunit}

Present neutrino experiments put stringent constraints on the deviations
from unitarity~\cite{Gluza:2002vs, Antusch:2006vwa,
FernandezMartinez:2007ms, Antusch:2014woa, Fernandez-Martinez:2016lgt,
Blennow:2016jkn}. In the framework of the type-I seesaw, it is the block $K$
of the matrix $\mathcal{V}$ that takes the role played by the $U_\text{PMNS}$
matrix at low energies, 
typically taken as unitary and parametrised accordingly
(see e.g.~the standard parametrisation~\cite{PDG2019}).
Clearly, in this framework, $K$ is no longer a unitary matrix.
When considering the deviations from unitarity of $K$, one must comply with
experimental bounds, while at the same time investigate whether it is
possible to obtain deviations that are
sizeable enough to be detected experimentally in the near future.
Using the above parametrisation,
this translates into making appropriate choices for the matrix $X$.
Deviations from unitarity of $K$ can be
parametrised as the product of an Hermitian matrix by a unitary matrix~\cite{Fernandez-Martinez:2016lgt}: 
\begin{align}
K\,=\,(\id-\eta )\,V \,,
\label{eq:khu}
\end{align}%
where $\eta$ is an Hermitian matrix. 
In the previous section,
we have instead parametrised $K$ with an Hermitian matrix to the right
and the unitary matrix $V$ to the left, see Eq.~\eqref{eq:epl}.
These right- and left-polar decompositions are unique
since we are dealing with a non-singular $K$ by assumption.
Moreover, they can be connected explicitly:
\begin{align}
\eta \,=\,
V\left(\id - \sqrt{\left(\id + X^\dagger X \right)^{-1}}\right)V^\dagger
\,=\,\id-U_K \left( \sqrt{\id +d_{X}^{2}}\ \right)^{-1}  U_K^\dagger\,.
\end{align}
Expanding in powers of $X$ (or equivalently of $d_X$), one obtains
\begin{align}
\eta 
\,=\, \frac{1}{2} \,U_K\, d_X^2 \,U_K^\dagger + \mathcal{O}(d_X^4)
\,=\, \frac{1}{2} \,V \, X^\dagger X\, V^\dagger + \mathcal{O}(X^4)\,.
\end{align}

Constraints on the entries of $\eta$ depend on the mass scale of the new neutrinos.
Bounds on $\eta$ can be found in the literature
for the scenario in which all three heavier neutrinos have
masses above the EW scale~\cite{Antusch:2014woa, Fernandez-Martinez:2016lgt}.
As pointed out in \cite{Fernandez-Martinez:2016lgt}, in such a case
it is very useful to parametrise $K$ with the
unitary matrix on the right, due to the fact that, experimentally, it is not
possible to determine which physical light neutrino is produced.
Therefore, one must sum over the massive neutrino fields
and observables depend on $KK^\dagger$.
From the unitarity relation $KK^\dagger+RR^\dagger=\id$
and Eq.~\eqref{eq:khu}, one has
\begin{align}
K K^\dagger \,=\, \id - R R^\dagger \,=\, \id - 2\eta + \eta^2
\quad \Rightarrow \quad
\eta \,=\, \frac{1}{2} \,R R^\dagger + \mathcal{O}(R^4)\,,
\label{eq:KReta}
\end{align}%
i.e.~there is a straightforward connection between $KK^{\dagger}$, $RR^{\dagger}$
and the deviations
from unitarity, expressed in $\eta$.

When one has one or more light sterile neutrinos,
the aforementioned bounds cannot be directly applied,
as some states are kinematically accessible and different sets of
experimental constraints need to be taken into account,
depending on the spectrum at hand.
In this case, observables can constrain directly the entries of $R$, and not just the product $RR^{\dagger}$.
For light sterile neutrinos with eV-scale masses, the most stringent bounds on deviations from unitarity
come from oscillation experiments~\cite{Blennow:2016jkn}, such as 
BUGEY-3~\cite{Declais:1994su}, MINOS~\cite{MINOS:2016viw}, NOMAD~\cite{Astier:2001yj,Astier:2003gs}
and Super\discretionary{-}{-}{-}Kamiokande~\cite{Abe:2014gda}.
In our analysis, the relevant exclusion curves in the $\sin^{2}2\vartheta_{\alpha \beta}$ -- $\Delta m^{2}$
planes (see section~\ref{sec:numeric}) are considered and translated into constraints
on the elements of the mixing matrix block $R$. 
If one is dealing instead with keV or GeV\,--\,TeV sterile neutrinos, it is 
important to take into account the experimental bounds coming
from $\beta$-decay experiments (see e.g.~\cite{Adhikari:2016bei} and references within) 
and from LHC searches for heavy Majorana neutrinos~\cite{Antusch:2015mia,
Deppisch:2015qwa, Das:2015toa, Das:2017nvm, Das:2017zjc,
Sirunyan:2018mtv, Sirunyan:2018xiv}.
Another crucial experimental input, also taken into account
in our analysis, is the limit on the $\mu \rightarrow e\gamma $ branching ratio
obtained by the MEG Collaboration, 
$BR(\mu \rightarrow e\gamma) < 4.2\times 10^{-13}$ ($90\%$~CL)~\cite{TheMEG:2016wtm},
one of the most stringent bounds on lepton flavour violating processes.
This bound is expected to be relevant whenever the heavier neutrino masses are
around or above the EW scale, as a GIM cancellation arises for lighter states
(see for instance Eq.~(40) of Ref.~\cite{Fernandez-Martinez:2016lgt}).

\subsection{Restrictions on the Neutrino Mass Spectrum}

The type-I seesaw model that we consider here, with at least one sterile
neutrino with a mass around 1 eV, also leads to some restrictions on the light
neutrino mass spectrum at tree level.
In particular, we find 
an upper bound on the mass $m_\text{min}$ of the lightest neutrino,
as a function of the deviations from unitarity.

Taking into account the parametrisation~\eqref{eq:xc} for the matrix $X$
controlling deviations from unitarity, and
for eigenvalues $d_{X_{i}}^2$ ($i=1,2,3$) of $X^\dagger X$, we have: 
\begin{align}
\tr\left[ X^{\dagger }X\right]\,=\,
\tr\left[ O_c^\dagger \, D^{-1}\, O_{c}\, d\right]
\,=\, d_{X_{1}}^{2}+d_{X_{2}}^{2}+d_{X_{3}}^{2}\,.  \label{ei}
\end{align}%
From this, and recalling that $d=\diag(m_1,m_2,m_3)$ and
$D=\diag(M_1, M_2, M_3)$, we obtain 
\begin{align}
\sum_k\,\frac{1}{M_k}
\left( m_{1}\left\vert O_{k1}^{c}\right\vert^{2}
      +m_{2}\left\vert O_{k2}^{c}\right\vert^{2}
      +m_{3}\left\vert O_{k3}^{c}\right\vert^{2}\right)
\,=\, d_{X_{1}}^{2}+d_{X_{2}}^{2}+d_{X_{3}}^{2}\,,
\label{ei1}
\end{align}
and conclude that%
\begin{align}
\frac{m_\text{min}}{M_{1}}\left( \left\vert O_{11}^{c}\right\vert ^{2}+\left\vert
O_{12}^{c}\right\vert ^{2}+\left\vert O_{13}^{c}\right\vert ^{2}\right)
\,<\,d_{X_{1}}^{2}+d_{X_{2}}^{2}+d_{X_{3}}^{2}\,,
\end{align}%
where naturally $M_{1}\leq M_{2}\leq M_{3}$ and $m_\text{min} = m_1$ ($m_3$) for normal (inverted) ordering.
Then, inserting the inequality $\sum_i \left\vert O_{1i}^{c}\right\vert
^{2}\geq 1$, valid for any orthogonal complex matrix, we find
\begin{align}
m_\text{min} \,<\, \left(d_{X_{1}}^{2}+d_{X_{2}}^{2}+d_{X_{3}}^{2}\right) M_{1} \,. \label{upm1}
\end{align}

As discussed, when one has one or more light sterile neutrinos, the typical
stringent conditions on the deviations from unitarity do not apply.
Thus, one may consider larger deviations from unitarity, even of the
order of the smallest $U_\text{PMNS}$ angle, i.e.~$\mathcal{O}(0.1)$~\cite{Blennow:2016jkn}.
Since in the scenarios of interest the lightest of the heaviest neutrinos
has a mass of $M_{1} \sim 1$ eV, using Eq.~\eqref{upm1} we
find a bound for the mass of the lightest neutrino:%
\begin{align}
m_\text{min} \,\lesssim \, 0.1 \text{ eV}\,.
\label{m1ev}
\end{align}
Note that this bound becomes stronger as one considers smaller
and smaller deviations from unitarity.
Taking into account the measured light neutrino mass-squared differences, we conclude
that the light neutrinos cannot have masses above $\mathcal{O}(0.1)$ eV
under these conditions, a statement which is also supported by cosmological bounds~%
\cite{Aghanim:2016yuo}.

\subsection{Neutrino Oscillations}
\label{sec:osc}

In the presence of deviations from unitarity,
neutrino oscillation probabilities are modified%
~\cite{Antusch:2006vwa,Blennow:2016jkn}.
If $n$ of the heavier neutrinos are accessible at oscillation experiments,
then a $3 \times (3 + n)$ submatrix $\Theta$ of $\mathcal{V}$ enters
the computation of oscillation probabilities,
\begin{align}
\Theta\,=\,
\left( \begin{array}{cc}
K & R_{3\times n}
\end{array} \right)\,,
\label{eq:wdef}
\end{align}
where $R_{3\times n}$ contains the first $n$ columns of $R$.
For a given experimental setup, and depending on their masses,
the heavier states may already be produced incoherently or instead
lose coherence before reaching the detector, due to wave-packet separation
(see e.g.~\cite{Cozzella:2018zwm}).
The probability of transition between flavour (anti-)neutrinos
\stackon[-.7pt]{$\nu$}{\brabar}$_\alpha$ and 
\stackon[-.7pt]{$\nu$}{\brabar}$_\beta$,
or of survival for a given flavour ($\alpha = \beta$),
with $\alpha, \beta = e, \mu, \tau$, can be shown to take the form
\begin{equation}
\begin{aligned}
P_{\stackon[-.7pt]{$\nu$}{\brabar}_\alpha \rightarrow \stackon[-.7pt]{$\nu$}{\brabar}_\beta}(L,E)
\,=\,\frac{1}{(\Theta\Theta^\dagger)_{\alpha\alpha}(\Theta\Theta^\dagger)_{\beta\beta}}
\Bigg[ 
\left|(\Theta\Theta^\dagger)_{\alpha\beta}\right|^2
&-   4 \sum_{i>j}^{3+n}\,\re
\left(\Theta_{\alpha i}^*\,\Theta_{\beta i}\,\Theta_{\alpha j}\,\Theta_{\beta j}^*\right)
\sin^2 \Delta_{ij} \\
&\pm 2 \sum_{i>j}^{3+n}\,\im
\left(\Theta_{\alpha i}^*\,\Theta_{\beta i}\,\Theta_{\alpha j}\,\Theta_{\beta j}^*\right)
\sin 2 \Delta_{ij}
\Bigg] \,,
\end{aligned}
\label{eq:probability}
\end{equation}%
where the plus or minus sign in the second line refers to neutrinos or anti-neutrinos, respectively.
Here, $L$ denotes the source-detector distance, $E$ is the (anti-)neutrino energy,
and one has defined
\begin{align}
\Delta_{ij} \,\equiv \, \frac{\Delta m^2_{ik}\, L}{4E}
\,\simeq\, 1.27\,\frac{\Delta m_{ij}^{2}[\text{eV}^{2}]\,L[\text{km}] }{ E[\text{GeV}]}\,,
\end{align}
with mass-squared differences $\Delta m_{ij}^2 \equiv m_i^2 - m_j^2$, as usual.

Note that if $n=3$
then $\Theta \Theta ^{\dagger}=KK^{\dagger}+RR^{\dagger}=\id_{3\times 3}$
due to the unitarity of the full $6\times 6$ mixing matrix $\mathcal{V}$
and Eq.~\eqref{eq:probability} reduces to the usual unitary formula.
It should be pointed out that the normalisation
$(\Theta\Theta^\dagger)_{\alpha\alpha}(\Theta\Theta^\dagger)_{\beta\beta}$
in \eqref{eq:probability} will cancel in the experimental event rates,
due to similar correction factors appearing in production rates
and detection cross-sections~\cite{Antusch:2006vwa,Cozzella:2018zwm}.
Nevertheless, we explicitly keep it in subsequent expressions.
It will turn out to be negligibly close to unity for our particular numerical examples.
The term proportional to $|(\Theta\Theta^\dagger)_{\alpha\beta}|^2$ is instead known to
correspond to a ``zero-distance'' effect~\cite{Langacker:1988ur,Antusch:2006vwa}.
It will also turn out to be negligible for our explicit numerical examples.

In what follows,
we will consider approximate forms of Eq.~\eqref{eq:probability},
having in mind SBL and long-baseline (LBL) experimental setups.
Since LBL experiments realistically need to take matter effects into account,
our formulae in those cases are simply indicative.

\section{Structure of the Mass Matrix}
\label{sec:loopandsym}
\subsection{One-loop Corrections}
\label{sec:loop}

So far we have focused on neutrino masses and mixing at tree level.
However, in general, one expects one-loop corrections $\delta M_{L}$
to the $0_{(3\times 3)}$ block of $\mathcal{M}$ in Eq.~\eqref{vm0}.
As these are not guaranteed to be negligible,
one should keep track of them in order to properly scan the parameter space of seesaw models.
They are inherently finite and are given by~\cite{Grimus:2002nk, AristizabalSierra:2011mn}
(see also \cite{Grimus:2018rte}): 
\begin{align}
\delta M_{L}\,=\,\delta M_{L}^{Z}+\delta M_{L}^{H}~, 
\end{align}
where $\delta M_{L}^{Z}$ and $\delta M_{L}^{H}$ represent contributions
depending on the $Z$ and Higgs boson masses, $m_Z$ and $m_H$, respectively. 
Explicitly, one has (see also Appendix A of Ref.~\cite{AristizabalSierra:2011mn}):
\begin{equation}
\begin{aligned}
\delta M_{L}^{Z} &\,=\,
\frac{3}{32\pi ^{2}\, v^2 }\,
\left( \begin{array}{cc} K & R \end{array} \right)\,
\frac{\mathcal{D}^{3}}{\mathcal{D}^{2}/{m_{Z}^{2}} - \id}\,
\log \left( \frac{\mathcal{D}^{2}}{m_{Z}^{2}} \right)\,
\left( \begin{array}{c} K^T \\ R^T \end{array} \right)\,, \\[2mm]
\delta M_{L}^{H} &\,=\,
\frac{1}{32\pi ^{2}\, v^2 }\,
\left( \begin{array}{cc} K & R \end{array} \right)\,
\frac{\mathcal{D}^{3}}{\mathcal{D}^{2}/{m_{H}^{2}} - \id}\,
\log \left( \frac{\mathcal{D}^{2}}{m_{H}^{2}} \right)\,
\left( \begin{array}{c} K^T \\ R^T \end{array} \right)\,, 
\label{dM1}
\end{aligned}
\end{equation}
in a generic weak basis, with $v \simeq 174$ GeV being the Higgs VEV
and with $\mathcal{D}$, $K$ and $R$ given in Eqs.~\eqref{eq:d} and \eqref{eq:epl}.
This result can be cast in a simple form:
\begin{align}
\delta M_{L}\,=\,
K\,f(d)\,K^T + R\,f(D)\,R^T\,, 
\label{eq:oneloop}
\end{align}
where naturally $f$ is applied element-wise to diagonal matrices,
with
\begin{align}
f(m) \,\equiv \, \frac{m^3}{(4\pi\,v)^2} \left(\frac{3\log(m/m_Z)}{m^2/m_Z^2 -1} + \frac{\log(m/m_H)}{m^2/m_H^2 -1}\right)\,.
\end{align}
Models with very small deviations from unitarity (standard seesaw) 
have a very small $X$ and hence a correspondingly small $R = K X^\dagger$.
For these, the one-loop $\delta M_{L}$ corrections are negligible,
as can be seen from Eq.~\eqref{eq:oneloop}. 
Namely (aside from the loop-factor suppression),
the terms with $K$ are suppressed by the light neutrino masses 
$d$, whereas the effect of the heavier neutrino masses in $D$ is regulated
by the small entries of $R$.
However, in models with sizeable deviations from unitarity,
$R$ is not small and controlling $\delta M_{L}$
requires a mechanism such as a symmetry at the Lagrangian level.

\subsection{Approximately Conserved Lepton Number}
\label{sec:approximateL}
Relatively light sterile neutrinos can arise naturally in a seesaw framework
in the presence of an approximately conserved lepton number~%
\cite{Shaposhnikov:2006nn, Kersten:2007vk, Ibarra:2010xw}.
Such a U(1)$_L$ symmetry, when exact, imposes specific textures on the mass matrices $m$ and $M$.
These textures may be slightly perturbed when the symmetry is approximate,%
\footnote{We allow for small perturbations to all entries of $m$ and $M$,
without any presumption regarding their origin.
The case where only $M$ departs from its symmetric texture,
which manifestly corresponds to a soft breaking of the lepton number symmetry,
was considered in Refs.~\cite{Lavoura:2000ci,Grimus:2001ex}.}
allowing for non-vanishing Majorana neutrino masses and non-trivial mixing.

We are interested in scenarios where at least one of the 
mostly-sterile neutrinos is light, with a mass of $\mathcal{O}$(eV),
in order to establish a connection to the SBL anomalies. 
We are further looking for situations where some of the Yukawa
couplings are of order one.
The choice of lepton charges should then be such that,
in the exact conservation limit:
i) $M$ has zero determinant,%
\footnote{In previous work \cite{Agostinho:2017wfs}, several cases were analysed following the U(1)$_L$
charge assignment $\lambda_\nu = (1,-1,0)$ and $\lambda_L=(1,1,1)$, which
however implies $\det \,M \neq 0$ in the symmetric limit.}
and ii) not all entries of $m$ are small.
These conditions limit the possible U(1)$_L$ charge assignments.

The possibility of having a conserved (non-standard) lepton number
has been considered in the past~\cite{Wolfenstein:1981kw,Leung:1983ti,Branco:1988ex}.
Following the analysis of Ref.~\cite{Branco:1988ex},
we work in a certain `symmetry' weak basis in which 
lepton charge vectors $\lambda_\nu$ and $\lambda_L$ 
are assigned to the three right-handed neutrino singlets
and to the three lepton doublets, respectively.
As anticipated in section~\ref{sec:param}, one generically has $\Sigma \neq \id$
in Eq.~\eqref{eq:Zgeneral}.
Up to permutations, there are only 4 non-trivial
choices of U(1)$_L$ charges leading to an $M$ with zero determinant
in the exact conservation limit:
$\lambda_\nu=(1,1,0)$, $\lambda_\nu=(1,-1,-1)$,
$\lambda_\nu=(1,1,1)$ and $\lambda_\nu=(0,0,1)$. 
Of these four, $\lambda_\nu=(1,1,1)$ is not viable as it imposes $M=0$,
and $\lambda_\nu=(0,0,1)$ is discarded since
requiring controlled loop corrections in our framework
effectively reduces it to the case with $\lambda_\nu = (1,-1,-1)$.
We look into in the remaining two options $\lambda_\nu=(1,1,0)$ and $\lambda_\nu=(1,-1,-1)$ in what follows.
Given $\lambda_\nu$, the choice of $\lambda_L$ follows
from the requirements that the seesaw mechanism is operative for all light neutrinos
and that all left-handed neutrinos are allowed to couple to
the right-handed ones~\cite{Branco:1988ex}.

\subsubsection{Case I: \texorpdfstring{$\lambda_\nu = (1,1,0)$}{lambda nu = (1,1,0)}}
\label{sec:caseI}
For this case, the only sensible choice for the doublet charges is
$\lambda_L = (0,0,0)$.
The mass matrices in the symmetric limit read:
\begin{align}
m\,=\,\left( 
\begin{array}{ccc}
0 & 0 & a \\ 
0 & 0 & b \\ 
0 & 0 & c%
\end{array}%
\right) \,,\quad
M\,=\,\left( 
\begin{array}{ccc}
0 & 0 & 0 \\ 
0 & 0 & 0 \\ 
0 & 0 & M_3%
\end{array}%
\right) \,.
\label{eq:MI}
\end{align}
Breaking the symmetry will generate the light neutrino masses, two
(mostly-)sterile states with masses $M_1$ and $M_2$ that can be much smaller than 
$M_3$, and a heavy sterile with a mass close to $M_3$. 
As expected, some Yukawa couplings remain of $\mathcal{O}(1)$,
which can also be understood from Eq.~\eqref{eq:mdirac}, expressing the
dependence of the Dirac mass matrix $m$ on the sterile masses contained in $D$.
This case is further separated into two subcases: one can allow for a hierarchy $M_2 \gg M_1$ ({\bf case Ia}),
which may arise in a scenario of stepwise symmetry breaking, or instead focus on a single new light-sterile scale,
with $M_1 \sim M_2$ ({\bf case Ib}).

\subsubsection{Case II: \texorpdfstring{$\lambda_\nu = (1,-1,-1)$}{lambda nu = (1,-1,-1)}}
\label{sec:caseII}
For this case, one is instead led to
$\lambda_L = (1,1,1)$. In the exact conservation limit,
the mass matrices are given by: 
\begin{align}
m\,=\,\left( 
\begin{array}{ccc}
a & 0 & 0 \\ 
b & 0 & 0 \\ 
c & 0 & 0%
\end{array}%
\right)\,,\quad
M\,=\,\left( 
\begin{array}{ccc}
0 & A & B \\ 
A & 0 & 0 \\ 
B & 0 & 0%
\end{array}%
\right)\,.
\label{eq:MII}
\end{align}
In this limit, one has two degenerate neutrinos with mass
$\sqrt{|A|^2+|B|^2}$ and opposite CP parities, forming a single heavy Dirac particle.
Breaking the symmetry will allow for the generation of light neutrino masses
and for another massive sterile state to arise, with a mass than can be much smaller than $|A|$ and $|B|$.
It will additionally lift the mass degeneracy for the Dirac neutrino,
producing a pseudo-Dirac neutrino pair~\cite{Wolfenstein:1981kw,Petcov:1982ya}.
As pointed out in~\cite{Ibarra:2011xn}, a strong mass degeneracy translates into a
symmetry in the $R$ block of the mixing matrix, namely $R_{\alpha 2}\simeq \pm i\, R_{\alpha 3}$ ($\alpha = e,\mu,\tau$).
Such a relation can be seen to play a fundamental role in suppressing the effect of the
large masses $M_2$ and $M_3$ in the one-loop correction $\delta M_L$, see Eq.~\eqref{eq:oneloop}. 
It signals that one is close to the limit of lepton number conservation,
even if $R_{\alpha 2}$ and $R_{\alpha 3}$ are not extremely suppressed.%
\footnote{Nonetheless, it is true that in the exact conservation limit $d=X=R=0$.}
One is then allowed to have relatively large Yukawa couplings
even if $M_2 \simeq M_3$ are not as large as the $M_3$ of case I.
This can be seen from Eq.~\eqref{eq:mdirac},
which can be written in the form $m \,=\, R\, D \left( Z^{-1}\right)^* $.
The mass of the pseudo-Dirac pair can be at the TeV scale~%
\cite{Ibarra:2010xw,Ibarra:2011xn,Dinh:2012bp,Cely:2012bz,Penedo:2017knr},
since the size of the lightest neutrino masses is protected by approximate lepton number conservation.
The same symmetry and effects are present in the examples given in Ref.~\cite{Agostinho:2017wfs}.

In the following section, we perform a numerical analysis focusing on cases Ia, Ib and II
and incorporating an eV sterile neutrino in the seesaw spectrum while allowing for a
mixing matrix $K$ with sizeable deviations from unitarity.

\section{Numerical Analysis and Benchmarks}
\label{sec:numeric}

For each of the cases Ia, Ib and II defined in the previous section,
we explicitly provide a numerical benchmark for the seesaw mass matrices,
and explore the parameter space of qualitatively similar seesaw structures.
As anticipated in section~\ref{sec:osc}, we further provide approximate forms
of the transition probabilities of muon to electron (anti-)neutrinos,
$P_{\stackon[-.7pt]{$\nu$}{\brabar}_\mu \rightarrow \stackon[-.7pt]{$\nu$}{\brabar}_e}$,
obtained from Eq.~\eqref{eq:probability} while having in mind SBL and LBL setups,
for each of the three scenarios.
Given that recent global fits~\cite{deSalas:2018bym,Esteban:2018azc} disfavour
a light neutrino mass spectrum with inverted ordering (IO) with respect to one with normal ordering (NO)
at more than the $3\sigma$ level, we restrict the mass ordering to NO in our numerical examples.

Before proceeding, note that the three scenarios of interest exhibit
some correspondence to the commonly considered
3+1+1 (case Ia), 3+2 (case Ib), and 3+1 (case II)
schemes, see for instance~\cite{Giunti:2019aiy}.
Thus, even though the connection to the latter is not exact --
in particular, the spectrum of case Ib is not that of a typical 3+2 scenario --
it may prove useful to consider quantities therein defined in our analysis,
namely~\cite{Gariazzo:2015rra}
\begin{align}
\sin^2 2 \vartheta^{(k)}_{\mu e}
\,\equiv\, 4\, \big|\Theta_{\mu k}\big|^2 \,\big|\Theta_{e k}\big|^2 
\,,
\label{eq:sthmue}
\end{align}
with $k=4$ in the 3+1 case, while $k=4,5$ for the other two cases.
According to the global fit to SBL data of Ref.~\cite{Gariazzo:2015rra},
explaining the observed anomalies requires
$\Delta m^2_{41} \in [0.87,\,2.04]$ eV$^2$ and
$\sin^2 2 \vartheta^{(4)}_{\mu e} \in [6.5\times 10^{-4},\, 2.6 \times 10^{-3}]$ ($99.7\%$ CL)
in the 3+1 scheme. This result may also be of relevance in the 3+1+1 scheme.
Although we take these intervals as guidelines in our numerical explorations,
it is not our aim to address the tensions in the current experimental situation of the SBL anomalies.
Thus, we only restrict our sterile neutrino parameter space at the outset through the conservative bounds
$\sum_i |R_{\alpha i}|^2 < 0.1$ ($\alpha = e,\mu,\tau$),
and via the constraints of \cite{Astier:2001yj, Astier:2003gs, Adhikari:2016bei}
on mixing matrix elements corresponding to
large mass-squared differences $\Delta m^2 \sim 10\text{ eV}^2 - 1\text{ keV}^2$,
as anticipated in section~\ref{sec:devunit}.

\subsection{Case Ia: \texorpdfstring{$M_1\ll M_2\ll M_3$}{M1 << M2 << M3}}

\begin{table}
\renewcommand{\thetable}{\arabic{table}a}
\centering
\renewcommand{\arraystretch}{1.2}
\begin{tabular}{lr}
\toprule
 & {\bf Case Ia} numerical benchmark \\
\midrule
\addlinespace
$m$ (GeV) &
$\begin{bmatrix*}[r]
 ( 2.11 -5.58\, i)\times 10^{-11} & ( 1.29 +1.65\, i)\times 10^{-9} &  11.2 -10.9\, i \\ 
 ( 0.85 +2.22\, i)\times 10^{-10} & (-5.29 +3.99\, i)\times 10^{-9} &  10.4 + 0.4\, i\\ 
 (-0.26 +1.98\, i)\times 10^{-10} & (-4.51 -1.05\, i)\times 10^{-9} & -10.5 -34.6\, i
\end{bmatrix*}$ \\
\addlinespace
$M$ (GeV)  & 
$\begin{bmatrix*}[r]
 8.93\times 10^{-10} & 4.45\times 10^{-11} & 1.28\times 10^{-13} \\
 4.45\times 10^{-11} & 1.00\times 10^{-6} & 6.22\times 10^{-11} \\
 1.28\times 10^{-13} & 6.22\times 10^{-11} & 5.00\times 10^{14} 
\end{bmatrix*}$
\\
\addlinespace
\midrule
\addlinespace
$K$  &
$\begin{bmatrix*}[r]
-0.797 +0.071 \, i &  0.578 +0.006 \, i & -0.115+0.096 \, i \\
 0.293 -0.086 \, i &  0.575 +0.027 \, i &  0.719+0.010 \, i \\
-0.516 -0.004 \, i & -0.570 +0.020 \, i &  0.606
\end{bmatrix*}$
\\
\addlinespace
$R$  &
$\begin{bmatrix*}[r]
 0.024 -0.057 \, i & ( 1.29 +1.65 \, i)\times 10^{-3} & (-2.24+2.18 \, i) \times 10^{-14} \\
 0.093 +0.223 \, i & (-5.29 +3.99 \, i)\times 10^{-3} & (-2.08+0.08 \, i) \times 10^{-14} \\
-0.026 +0.199 \, i & (-4.51 -1.05 \, i)\times 10^{-3} & (-2.10+6.92 \, i) \times 10^{-14}
\end{bmatrix*}$
\\
\addlinespace
$X$  &
$\begin{bmatrix*}[r]
-0.003-0.015 \, i & 0.102 +0.023 \, i & 0.050 -0.317 \, i \\
(-5.12+1.72 \, i)\times 10^{-4}  & ( 0.46 -4.33 \, i)\times 10^{-3}  & (-7.30-2.18 \, i)\times 10^{-3} \\
( 0.23+5.33 \, i)\times 10^{-14} & (-3.44 +2.75 \, i)\times 10^{-14} & ( 0.36-4.41 \, i)\times 10^{-14}
\end{bmatrix*}$
\\
\addlinespace
$O_c$ (tree level) \!\!\!\!\!\! \!\!\!\!\!\!\!\!\!\!\!\!& 
$\begin{bmatrix*}[r]
-0.53 +0.12 \, i &  0.22 -1.12 \, i & -1.41 -0.22 \, i \\
 0.22 +0.56 \, i & -1.50 -0.13 \, i & -0.30 +1.03 \, i \\
 1.00 -0.06 \, i &  0.23 +0.25 \, i & -0.14 -0.01 \, i
\end{bmatrix*}$
\\
\addlinespace
\midrule
\addlinespace
Masses &
$\begin{matrix*}[l]
m_1 \simeq 1.06\times 10^{-3}\text{ eV}\,,\quad & m_2 \simeq 8.48\times 10^{-3}\text{ eV} \,,\quad & m_3 \simeq 5.02\times 10^{-2}\text{ eV} \,,\,\\ 
M_1 \simeq 1.00\text{ eV}\,,\,               & M_2 \simeq 1.00\text{ keV} \,,\,              & M_3 \simeq 5.00\times 10^{14}\text{ GeV}
\end{matrix*}$ \\
\addlinespace
\midrule
\addlinespace
$3\nu$ $\Delta m^2$ &
$\Delta m^2_\odot = \Delta m^2_{21} \simeq 7.08 \times 10^{-5}\text{ eV}^2\,,\,
\quad\,\,\,\,
\Delta m^2_\text{atm} = \Delta m^2_{31} \simeq 2.52 \times 10^{-3}\text{ eV}^2$
\\
\addlinespace
$3\nu$ mixing angles \!\!\!\!\!\!\!\!\!\!\!\!& 
$\sin^2 \theta_{12} \simeq 0.344\,,\,\quad\,\,\,\,
\sin^2 \theta_{23} \simeq 0.585\,,\,\quad\,\,\,\,
\sin^2 \theta_{13} \simeq 0.0236$
\\
\addlinespace
$3\nu$ CPV phases \!\!\!\!\!\! & 
$\delta \simeq 1.21 \pi\,,\,\quad\,\,\,\,
\alpha_{21} \simeq 0.06 \pi\,,\,\quad\,\,\,\,
\alpha_{31} \simeq 0.06 \pi$
\\
\addlinespace
\midrule
\addlinespace
$\sin^2 2 \vartheta^{(i)}_{\mu e}$ & 
$\sin^2 2 \vartheta_{\mu e}^{(4)} \simeq 8.8\times 10^{-4}\,,\,\quad\,\,\,\,
\sin^2 2 \vartheta_{\mu e}^{(5)} \simeq 7.7\times 10^{-10}$
\\
\bottomrule
\end{tabular}
\caption{Numerical benchmark for case Ia. The ordering of light neutrinos is NO.
From the input matrices $m$ and $M$, and taking into account one-loop corrections, the other quantities here listed follow.
It should be noted that $O_c$ of Eq.~\eqref{eq:xc} is only defined at tree level.
Values for the mixing angles and CPV phases of the $3\nu$-framework in the standard parametrisation~\cite{PDG2019}
are extracted by identifying the unitary matrix $V$ with a unitary $3\times 3$ PMNS mixing matrix.}
\label{tab:Ia}
\end{table}
The numerical data for the benchmark corresponding to this case is given in Table~\ref{tab:Ia},
where the one-loop correction of Eq.~\eqref{eq:oneloop} has been taken into account.
Apart from the three light mostly\discretionary{-}{-}{-}active neutrinos,
the spectrum includes three mostly-sterile neutrinos with masses
$M_1 \sim 1$ eV, $M_2 \sim 1$ keV,
and $M_3$ a few orders of magnitude below the grand unification (GUT) scale, $M_3 \sim 10^{14}$ GeV.
The keV-scale neutrino may be a viable dark matter candidate~\cite{Adhikari:2016bei, Boyarsky:2018tvu}.

For the spectrum of case Ia, one has $n=2$ in Eq.~\eqref{eq:wdef}.
In the context of a LBL experiment (e.g.~DUNE~\cite{DUNE}),
the expression of Eq.~\eqref{eq:probability}
applied to the transition probability of muon to electron (anti-)neutrinos
can, in this case, be approximated by:
\begin{equation}
\begin{aligned}
P^\text{LBL}_{\stackon[-.7pt]{$\nu$}{\brabar}_\mu \rightarrow \stackon[-.7pt]{$\nu$}{\brabar}_e}
\,\simeq\,
&\frac{1}{(\Theta\Theta^\dagger)_{\mu\mu}(\Theta\Theta^\dagger)_{ee}}
\Bigg[ 
\left|(\Theta\Theta^\dagger)_{\mu e}\right|^2
\\ &
-   4 \sum_{i>j}^3\,\re
\left(\Theta_{\mu i}^*\,\Theta_{e i}\,\Theta_{\mu j}\,\Theta_{e j}^*\right)
\sin^2 \Delta_{ij} 
\pm 2 \sum_{i>j}^{3}\,\im
\left(\Theta_{\mu i}^*\,\Theta_{e i}\,\Theta_{\mu j}\,\Theta_{e j}^*\right)
\sin 2 \Delta_{ij}
\\ &
-   4 \,\cdot\,\frac{1}{2}\,\re
\left(\Theta_{\mu 4}^*\,\Theta_{e 4}\,\sum_{j=1}^3\,\Theta_{\mu j}\,\Theta_{e j}^*\right)
-   4 \,\cdot\,\frac{1}{2}\,\re
\left(\Theta_{\mu 5}^*\,\Theta_{e 5}\,\sum_{j=1}^4\,\Theta_{\mu j}\,\Theta_{e j}^*\right)
\Bigg] \,,
\label{eq:LBL1Ia}
\end{aligned}
\end{equation}
where terms depending on $\Delta_{4j},\,\Delta_{5j} \gg 1$ 
have been replaced by their averaged versions
($\sin^2 \Delta_{ij} \to 1/2$, $\sin 2\Delta_{ij} \to 0$).
%
%
While the normalisation and the first term in this equation signal the loss of unitarity and 
a zero-distance effect, respectively,
the last two terms explicitly represent the effects
of the two lightest mostly-sterile states in oscillations.
If one is in a condition similar to that of
the numerical benchmark of Table~\ref{tab:Ia},
for which $|(\Theta\Theta^\dagger)_{\mu\mu}(\Theta\Theta^\dagger)_{ee} - 1|$
and $|(\Theta\Theta^\dagger)_{\mu e}|^2$ are negligible,
this expression can be further approximated by:
\begin{align}
P^\text{LBL}_{\stackon[-.7pt]{$\nu$}{\brabar}_\mu \rightarrow \stackon[-.7pt]{$\nu$}{\brabar}_e}
\,\simeq\,
P^{\text{LBL, }3\nu}_{\stackon[-.7pt]{$\nu$}{\brabar}_\mu \rightarrow \stackon[-.7pt]{$\nu$}{\brabar}_e}
+ \frac{1}{2}\sin^2 2 \vartheta^{(4)}_{\mu e}
 \,,
\label{eq:LBL2Ia}
\end{align}
where we have defined a $3\nu$-framework transition probability which, however, incorporates the effects of deviations of $K$ from unitarity,
\begin{align}
P^{\text{LBL, }3\nu}_{\stackon[-.7pt]{$\nu$}{\brabar}_\mu \rightarrow \stackon[-.7pt]{$\nu$}{\brabar}_e}
\,\equiv\,
-   4 \sum_{i>j}^3\,\re
\left(\Theta_{\mu i}^*\,\Theta_{e i}\,\Theta_{\mu j}\,\Theta_{e j}^*\right)
\sin^2 \Delta_{ij} 
\pm 2 \sum_{i>j}^{3}\,\im
\left(\Theta_{\mu i}^*\,\Theta_{e i}\,\Theta_{\mu j}\,\Theta_{e j}^*\right)
\sin 2 \Delta_{ij}
\,,
\end{align}
and have used the definition of Eq.~\eqref{eq:sthmue},
the unitarity of the full $6\times 6$ mixing matrix, and the fact that
$|\Theta_{\alpha 4}|^2 (= |R_{\alpha 1}|^2) \gg |\Theta_{\alpha 5}|^2 (= |R_{\alpha 2}|^2) \ggg |R_{\alpha 3}|^2$.

%
In a SBL experiment (e.g.~MicroBooNE~\cite{MicroBooNE}),
the relevant form of Eq.~\eqref{eq:probability} for
$\stackon[-.7pt]{$\nu$}{\brabar}_\mu \rightarrow \stackon[-.7pt]{$\nu$}{\brabar}_e$
transitions is:
\begin{equation}
\begin{aligned}
P^\text{SBL}_{\stackon[-.7pt]{$\nu$}{\brabar}_\mu \rightarrow \stackon[-.7pt]{$\nu$}{\brabar}_e}
\,\simeq\,
&\frac{1}{(\Theta\Theta^\dagger)_{\mu\mu}(\Theta\Theta^\dagger)_{ee}}
\Bigg[ 
\left|(\Theta\Theta^\dagger)_{\mu e}\right|^2
-   4 \,\cdot\,\frac{1}{2}\,\re
\left(\Theta_{\mu 5}^*\,\Theta_{e 5}\,\sum_{j=1}^4\,\Theta_{\mu j}\,\Theta_{e j}^*\right)
\\ &
-   4 \,\re
\left(\Theta_{\mu 4}^*\,\Theta_{e 4}\,\sum_{j=1}^3\,\Theta_{\mu j}\,\Theta_{e j}^*\right)
\sin^2 \Delta_{41} 
\pm 2 \,\im
\left(\Theta_{\mu 4}^*\,\Theta_{e 4}\,\sum_{j=1}^{3}\,\Theta_{\mu j}\,\Theta_{e j}^*\right)
\sin 2 \Delta_{41}
\Bigg] \,,
\label{eq:SBL1Ia}
\end{aligned}
\end{equation}
with $\Delta_{41}\simeq \Delta_{42}\simeq \Delta_{43}$, and
where terms depending on $\Delta_{5j} \gg 1$ 
have been replaced by their averaged versions
($\sin^2 \Delta_{5j} \to 1/2$, $\sin 2\Delta_{5j} \to 0$).
In this context, one is sensitive to oscillations due to the scale of the mass-squared differences $\Delta m^2_{4j}$ with $j=1,2,3$,
while the oscillations pertaining to smaller mass-squared differences have not yet had a chance to develop.
%
%
Finally, if one is in a condition similar to that of
the numerical benchmark, this expression can be simply approximated by:
\begin{align}
P^\text{SBL}_{\stackon[-.7pt]{$\nu$}{\brabar}_\mu \rightarrow \stackon[-.7pt]{$\nu$}{\brabar}_e}
\,\simeq\, \sin^2 2 \vartheta^{(4)}_{\mu e}\,\sin^2 \Delta_{41} \,,
\label{eq:SBL2Ia}
\end{align}
where once again one has taken into account
the unitarity of the full mixing matrix and the fact that
$|R_{\alpha 1}|^2 \gg  |R_{\alpha 2}|^2 \ggg |R_{\alpha 3}|^2$.

\begin{figure}[t]
\centering
\includegraphics[width=1.0\linewidth]{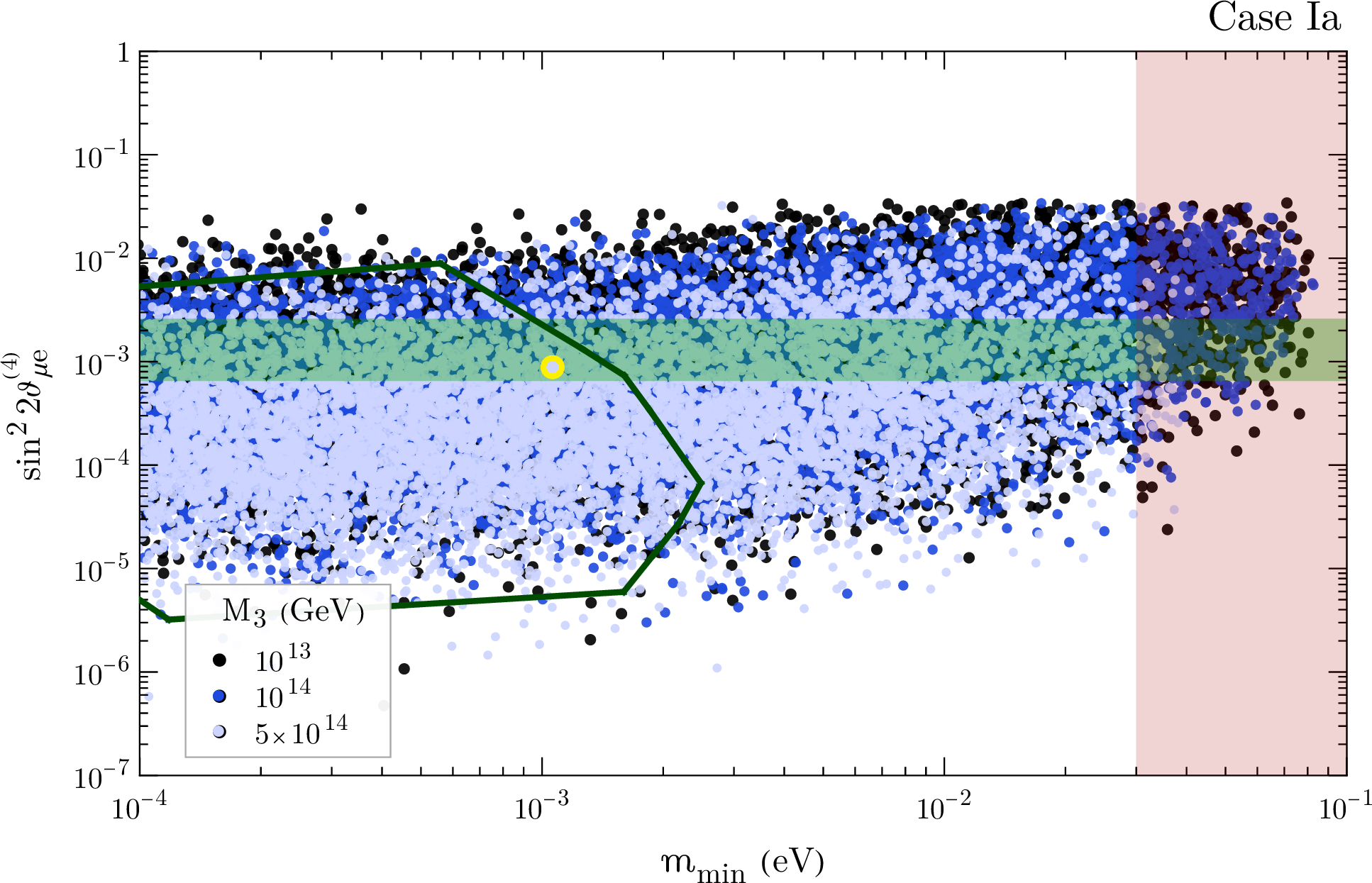}
\caption{
Active-sterile mixing measure $\sin^2 2\vartheta_{\mu e}^{(4)}$ 
{\it versus} the lightest neutrino mass $m_\text{min}$ from a scan of the case-Ia parameter space,
with NO ($m_\text{min} = m_1$). The heavy spectrum at tree level has
$M_1 = 1$ eV and $M_2 = 1$ keV, while three values of
the heaviest mass are considered, $M_3 = 10^{13}\, (10^{14})\, [5\times 10^{14}]$ GeV,
corresponding to the black (dark blue) [light blue] points in the scatter plot.
The horizontal green band shows the $99.7\%$ CL interval of Ref.~\cite{Gariazzo:2015rra},
while the vertical red exclusion band is obtained by combining
the most stringent bound on the sum of light neutrino masses from cosmology,
$\sum_i m_i < 0.12$ eV ($95\%$ CL)~\cite{Vagnozzi:2017ovm}\cite{Aghanim:2018eyx},
with the $3\sigma$ ranges of mass-squared differences.
The dark green contour delimits the region inside which loop-stable points have been found (see text),
while the benchmark of Table~\ref{tab:Ia} is marked in yellow.}
\label{fig:Ia}
\end{figure}
To further explore the parameter space of case Ia, we have
produced numerical seesaw structures by specifying
tree-level values of the unitary part $V$ of the mixing matrix $K$,
the mostly-active and mostly-sterile masses in $d$ and $D$,
and by scanning the complex orthogonal matrix $O_c$, parametrised
as a product of three complex rotations times a sign corresponding to its determinant.
We are interested in seesaw structures qualitatively similar to our benchmark,
so that we specify (at tree level) $M_1 = 1$ eV and $M_2 = 1$ keV,
while considering three different values for the heaviest neutrino mass,
$M_3 \in \{10^{13},\, 10^{14},\, 5\times 10^{14}\}$ GeV. 
While the lightest neutrino mass $m_\text{min}$ is scanned
in the range $[10^{-4},\,0.1]$ eV, the remaining elements of $d$ are fixed 
by specifying the solar and atmospheric mass differences.
The $3\nu$ mixing angles and Dirac CPV phase entering $V$ as well as
the aforementioned $3\nu$ mass-squared differences
are chosen to be the central values of the global fit of Ref.~\cite{Esteban:2018azc}.
We stress that, as was the case for the numerical benchmark of Table~\ref{tab:Ia},
$3\nu$ mixing angles and CPV phases obtained while
identifying $V$ with a unitary $3\times 3$ mixing matrix
are expected to deviate slightly from the mixing angles and CPV phases arising in
a parametrisation of the full $6\times 6$ mixing matrix $\mathcal{V}$,
due to deviations from unitarity.
In Figure~\ref{fig:Ia}
we show the values of $\sin^2 2\vartheta_{\mu e}^{(4)}$ in Eq.~\eqref{eq:sthmue}
against the values of the lightest neutrino mass, for the numerical examples found for case Ia.
Only points for which $\tr\left[m\,m^\dagger\right] \in [0.01,\,1]\,v^2$ are kept.%
\footnote{One may avoid very small Yukawa couplings by choosing appropriate values for $M_3$ and $O_c$.}
The horizontal green band highlights the range of $\sin^2 2\vartheta_{\mu e}^{(4)}$
preferred by the global fit of Ref.~\cite{Gariazzo:2015rra} and cited at the beginning of this section.
The dark green contour instead delimits the region inside which relatively loop-stable points can be found,
i.e.~points which,
after the one-loop correction of Eq.~\eqref{eq:oneloop} has been implemented,
still have $3\nu$ mass-squared differences and mixing angles (extracted from $V$)
inside the $3\sigma$ ranges of the fit~\cite{Esteban:2018azc}.
From the figure it can be seen that
raising the scale of $M_3$ will lower
the scale of the light neutrino masses,
disallowing too large values of $m_\text{min}$.
The approximations used in deriving the oscillation formulae
of Eqs.~\eqref{eq:LBL2Ia} and~\eqref{eq:SBL2Ia}
hold for all the plotted points.

Some quantities of potential phenomenological relevance, unrelated to 
neutrino oscillations, include 
the effective electron neutrino mass in $\beta$-decay, $m_\beta$,
the absolute value of the effective neutrino Majorana mass controlling the rate
of neutrinoless double beta ($(\beta\beta)_{0\nu}$-)decay, $|m_{\beta\beta}|$,
and the $\mu \to e \gamma$ branching ratio, $BR(\mu \rightarrow e\gamma)$.
For all numerical examples pertaining to case Ia which are stable under loop corrections,
the latter is unobservably small $BR(\mu \rightarrow e\gamma) \ll 10^{-30}$,
while the former two are bounded by $m_\beta < 9.4$ meV and $|m_{\beta\beta}| < 6.7$ meV,
and hence still out of reach of present and near-future experiments.
In the computation of $|m_{\beta\beta}|$, the effects of the eV- and keV-scale
neutrinos have been taken into account.

In the presence of a relatively large active-sterile mixing, 
future KATRIN-like experiments may be sensitive
to the existence of sterile neutrinos with $\mathcal{O}$(eV) masses~\cite{Riis:2010zm}.
This sensitivity is controlled by $|R_{e1}|^2 = |\mathcal{V}_{e4}|^2$, which is found to be bounded by $|R_{e4}|^2 \lesssim 0.02$
for the loop\discretionary{-}{-}{-}stable numerical examples of this case.
Sterile neutrinos with $\mathcal{O}$(keV) masses may instead be detectable via kink-like signatures
in next-generation $\beta$-decay experiments,
even in the presence of small mixing $|R_{e2}|^2= |\mathcal{V}_{e5}|^2 \sim 10^{-6}$~\cite{Mertens:2014nha}.

\subsection{Case Ib: \texorpdfstring{$M_1\sim M_2\ll M_3$}{M1 \textasciitilde{} M2 << M3}}
\begin{table}
\addtocounter{table}{-1}
\renewcommand{\thetable}{\arabic{table}b}
\centering
\renewcommand{\arraystretch}{1.2}
\begin{tabular}{lr}
\toprule
 & {\bf Case Ib} numerical benchmark \\
\midrule
\addlinespace
$m$ (GeV) &
$\begin{bmatrix*}[r]
 ( 0.46 - 2.57\, i) \times 10^{-10} & ( 2.37 + 0.54\, i) \times 10^{-10} & 11.24 - 2.72\, i \\
 (-5.50 - 1.04\, i) \times 10^{-10} & ( 0.68 - 6.20\, i) \times 10^{-10} &  8.90 -27.50\, i \\
 (-3.69 + 1.78\, i) \times 10^{-10} & (-1.60 - 4.45\, i) \times 10^{-10} & -1.85 + 0.43\, i\end{bmatrix*}$ \\
\addlinespace
$M$ (GeV)  & 
$\begin{bmatrix*}[r]
 2.88\times 10^{-9}  & 8.24\times 10^{-11} & 1.41\times 10^{-11}\\
 8.24\times 10^{-11} & 2.87\times 10^{-9}  & 1.42\times 10^{-11}\\
 1.41\times 10^{-11} & 1.42\times 10^{-11} & 1.00\times 10^{14}
\end{bmatrix*}$
\\
\addlinespace
\midrule
\addlinespace
$K$  &
$\begin{bmatrix*}[r]
-0.799 +0.137 \, i &  0.558 +0.001 \, i &  0.116 -0.071 \, i \\
 0.272 -0.172 \, i &  0.582 -0.036 \, i & -0.695 +0.014 \, i \\
-0.480 +0.099 \, i & -0.560 +0.141 \, i & -0.620 -0.019 \, i
\end{bmatrix*}$
\\
\addlinespace
$R$  &
$\begin{bmatrix*}[r]
0.039 +0.077\, i &  0.067 -0.040\, i & (-1.12 +0.27\, i)\times 10^{-13} \\
0.156 -0.105\, i & -0.097 -0.170\, i & (-0.89 +2.75\, i)\times 10^{-13} \\
0.061 -0.140\, i & -0.115 -0.071\, i & ( 1.85 -0.43\, i)\times 10^{-14}
\end{bmatrix*}$
\\
\addlinespace
$X$  &
$\begin{bmatrix*}[r]
-0.003 +0.009\, i & 0.073 -0.064 \, i & -0.168 -0.196 \, i \\
-0.009 -0.005\, i & 0.049 +0.078 \, i &  0.170 -0.185 \, i \\
(1.40 -5.37 \, i)\times 10^{-14} & (-1.47 -1.74 \, i)\times 10^{-13} & (0.37 +2.24 \, i)\times 10^{-13}
\end{bmatrix*}$
\\
\addlinespace
$O_c$ (tree level) \!\!\!\!\!\! \!\!\!\!\!\!\!\!\!\!\!\!& 
$\begin{bmatrix*}[r]
-1.06 -0.51 \, i & -1.10 -1.29 \, i & -1.51 +1.30 \, i \\
 0.75 -1.08 \, i &  1.40 -0.83 \, i & -1.46 -1.35 \, i \\
 0.91 +0.31 \, i & -0.60 +0.44 \, i &  0.32 -0.05 \, i
\end{bmatrix*}$
\\
\addlinespace
\midrule
\addlinespace
Masses &
$\begin{matrix*}[l]
m_1 \simeq 0.24\times 10^{-3}\text{ eV}\,,\quad & m_2 \simeq 8.76\times 10^{-3}\text{ eV} \,,\quad & m_3 \simeq 5.00\times 10^{-2}\text{ eV} \,,\,\\ 
M_1 \simeq 3.00\text{ eV}\,,\,               & M_2 \simeq 3.16\text{ eV} \,,\,              & M_3 \simeq 1.00\times 10^{14}\text{ GeV}
\end{matrix*}$ \\
\addlinespace
\midrule
\addlinespace
$3\nu$ $\Delta m^2$ &
$\Delta m^2_\odot = \Delta m^2_{21} \simeq 7.66 \times 10^{-5}\text{ eV}^2\,,\,
\quad\,\,\,\,
\Delta m^2_\text{atm} = \Delta m^2_{31} \simeq 2.50 \times 10^{-3}\text{ eV}^2$
\\
\addlinespace
$3\nu$ mixing angles \!\!\!\!\!\!\!\!\!\!\!\!& 
$\sin^2 \theta_{12} \simeq 0.327\,,\,\quad\,\,\,\,
 \sin^2 \theta_{23} \simeq 0.562\,,\,\quad\,\,\,\,
 \sin^2 \theta_{13} \simeq 0.0232$
\\
\addlinespace
$3\nu$ CPV phases \!\!\!\!\!\! & 
$\delta \simeq 1.26 \pi\,,\,\quad\,\,\,\,
\alpha_{21} \simeq 0.11 \pi\,,\,\quad\,\,\,\,
\alpha_{31} \simeq 0.22 \pi$
\\
\addlinespace
\midrule
\addlinespace
$\sin^2 2 \vartheta^{(i)}_{\mu e}$ & 
$\sin^2 2 \vartheta_{\mu e}^{(4)} \simeq 1.1\times 10^{-3}\,,\,\quad\,\,\,\,
 \sin^2 2 \vartheta_{\mu e}^{(5)} \simeq 9.2\times 10^{-4}$
\\
\bottomrule
\end{tabular}
\caption{The same as Table~\ref{tab:Ia} for case Ib.}
\label{tab:Ib}
\end{table}
The numerical data for the benchmark corresponding to this case is given in Table~\ref{tab:Ib}.
Apart from the three light mostly-active neutrinos,
the spectrum includes three mostly-sterile neutrinos with masses
$M_1 \sim M_2 \sim 3$ eV, such that $M_2^2 - M_1^2 \simeq 1$ eV$^2$,
while $M_3 \sim 10^{14}$ GeV.

For the spectrum of case Ib, one has $n=2$ in Eq.~\eqref{eq:wdef}.
In a LBL context, the expression of Eq.~\eqref{eq:probability}
applied to the transition probability of muon to electron (anti-)neutrinos
can be approximated by the same expression~\eqref{eq:LBL1Ia} given for case Ia.
%
%
Once again,
the last two terms in that equation explicitly show the effects
of the two lightest mostly-sterile states in oscillations.
If one is in a condition similar to that of
the benchmark of Table~\ref{tab:Ib},
for which $|(\Theta\Theta^\dagger)_{\mu\mu}(\Theta\Theta^\dagger)_{ee} - 1|$
and $|(\Theta\Theta^\dagger)_{\mu e}|^2$ are negligible,
this expression can be further approximated by:
\begin{align}
P^\text{LBL}_{\stackon[-.7pt]{$\nu$}{\brabar}_\mu \rightarrow \stackon[-.7pt]{$\nu$}{\brabar}_e}
\,\simeq\,
P^{\text{LBL, }3\nu}_{\stackon[-.7pt]{$\nu$}{\brabar}_\mu \rightarrow \stackon[-.7pt]{$\nu$}{\brabar}_e}
\,+\, \frac{1}{2}
\Big[
 \sin^2 2 \vartheta^{(4)}_{\mu e}
+\sin^2 2 \vartheta^{(5)}_{\mu e}
+4\, \re
\left(\Theta_{\mu 4}^*\,\Theta_{e 4}\,\Theta_{\mu 5}\,\Theta_{e 5}^*\right)
\Big] \,,
\label{eq:LBL2Ib}
\end{align}
where we have used the unitarity of the full $6\times 6$ mixing matrix,
and the fact that $|R_{\alpha 1}|^2  \sim |R_{\alpha 2}|^2 \ggg |R_{\alpha 3}|^2$.
The latter prevents us from neglecting $|R_{\alpha 2}|^2$ (and hence $\sin^2 2\vartheta^{(5)}_{\mu e}$)
with respect to $|R_{\alpha 1}|^2$ (and $\sin^2 2\vartheta^{(4)}_{\mu e}$), as we did in the previous case.

%
In a SBL context,
the relevant form of Eq.~\eqref{eq:probability} for
$\stackon[-.7pt]{$\nu$}{\brabar}_\mu \rightarrow \stackon[-.7pt]{$\nu$}{\brabar}_e$
transitions in case Ib is:
\begin{equation}
\begin{aligned}
P^\text{SBL}_{\stackon[-.7pt]{$\nu$}{\brabar}_\mu \rightarrow \stackon[-.7pt]{$\nu$}{\brabar}_e}
\,\simeq\,
&\frac{1}{(\Theta\Theta^\dagger)_{\mu\mu}(\Theta\Theta^\dagger)_{ee}}
\Bigg[ 
\left|(\Theta\Theta^\dagger)_{\mu e}\right|^2
\\ &
-   4 \,\cdot\, \frac{1}{2} \,\re
\left(\Theta_{\mu 4}^*\,\Theta_{e 4}\,\sum_{j=1}^3\,\Theta_{\mu j}\,\Theta_{e j}^*\right)
-   4 \,\cdot\, \frac{1}{2} \,\re
\left(\Theta_{\mu 5}^*\,\Theta_{e 5}\,\sum_{j=1}^3\,\Theta_{\mu j}\,\Theta_{e j}^*\right)
\\ &
-   4 \,\re
\left(\Theta_{\mu 5}^*\,\Theta_{e 5}\,\Theta_{\mu 4}\,\Theta_{e 4}^*\right)
\sin^2 \Delta_{54} 
\pm 2 \,\im
\left(\Theta_{\mu 5}^*\,\Theta_{e 5}\,\Theta_{\mu 4}\,\Theta_{e 4}^*\right)
\sin 2 \Delta_{54}
\Bigg] \,,
\label{eq:SBL1Ib}
\end{aligned}
\end{equation}
where terms depending on the large $\Delta_{4j}$ and $\Delta_{5j}$ ($j=1,2,3$)
have been replaced by their averaged versions.
It is clear that this case does not correspond to a typical 3+2 scenario
(see for instance~\cite{Gariazzo:2015rra}),
since one has $\Delta m^2_{4j},\,\Delta m^2_{5j} \sim 10$ eV$^2$ for $j=1,2,3$.
Hence, one can be sensitive to oscillations due to the mass-squared difference $\Delta m^2_{54} \sim 1$ eV$^2$,
while oscillations pertaining to larger differences are averaged out and
those driven by smaller mass-squared differences are underdeveloped.
%
%
If one is in a condition similar to that of
the numerical benchmark, this expression can be approximated by:
\begin{equation}
\begin{aligned}
P^\text{SBL}_{\stackon[-.7pt]{$\nu$}{\brabar}_\mu \rightarrow \stackon[-.7pt]{$\nu$}{\brabar}_e}
\,\simeq\, 
 \frac{1}{2} \Big( \sin^2 2 \vartheta^{(4)}_{\mu e} +\sin^2 2 \vartheta^{(5)}_{\mu e} \Big) 
&+   4 \,\re\left(\Theta_{\mu 4}^*\,\Theta_{e 4}\,\Theta_{\mu 5}\,\Theta_{e 5}^*\right)\cos^2 \Delta_{54}\\
&\mp 2 \,\im\left(\Theta_{\mu 4}^*\,\Theta_{e 4}\,\Theta_{\mu 5}\,\Theta_{e 5}^*\right)\sin 2 \Delta_{54}
\,,
\label{eq:SBL2Ib}
\end{aligned}
\end{equation}
where once again we have taken into account
the unitarity of the full mixing matrix and the fact that
$|R_{\alpha 1}|^2 \sim  |R_{\alpha 2}|^2 \ggg |R_{\alpha 3}|^2$.
Notice that, unlike the typical 3+2 case,
oscillations here depend on the square of the cosine of the relevant $\Delta_{ij}$.

\begin{figure}[t]
\centering
\includegraphics[width=1.0\linewidth]{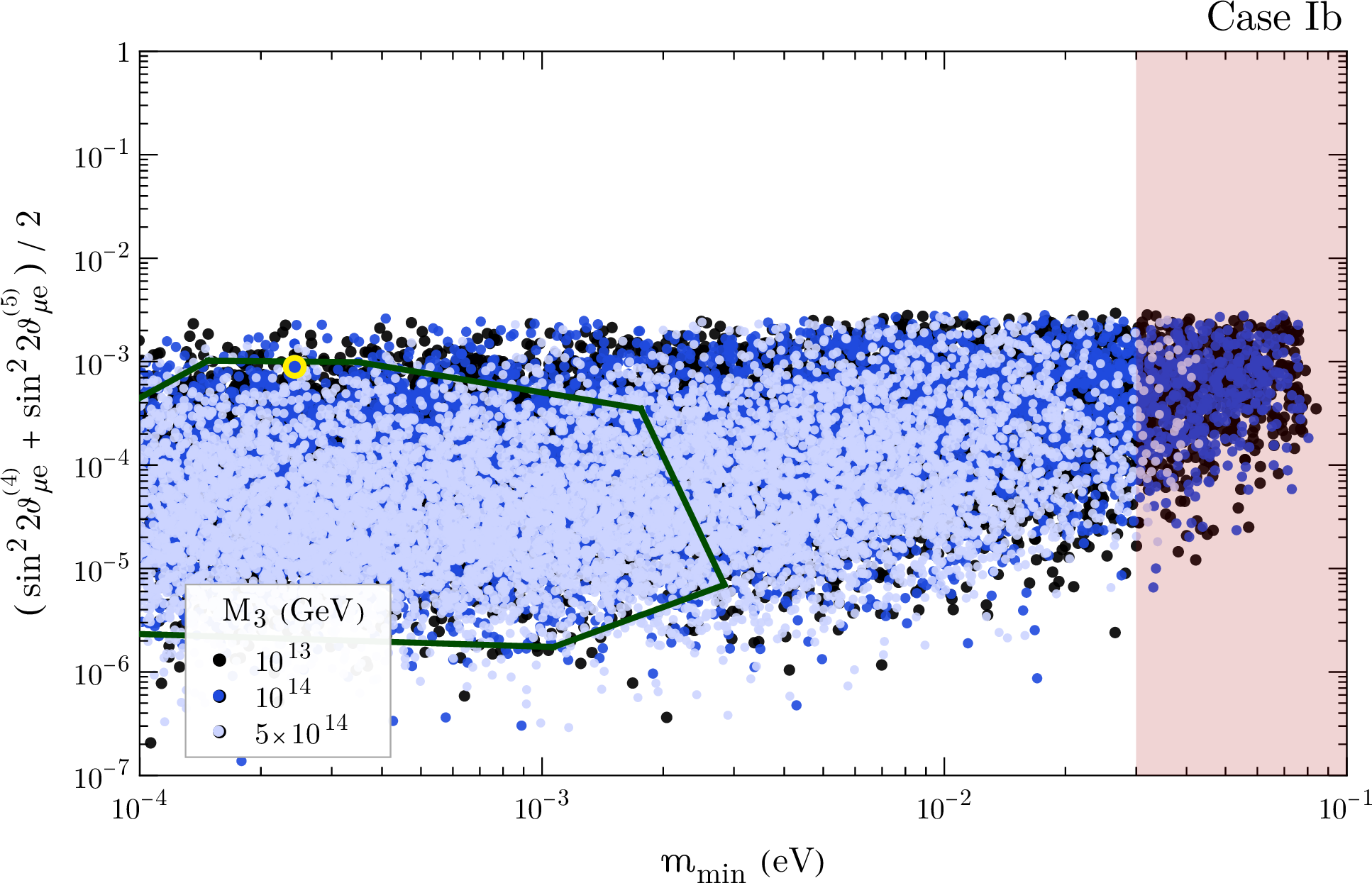}
\caption{The average of $\sin^2 2\vartheta_{\mu e}^{(4)}$ and $\sin^2 2\vartheta_{\mu e}^{(5)}$ 
{\it versus} the lightest neutrino mass $m_\text{min}$, from a scan of the case-Ib parameter space,
with NO ($m_\text{min} = m_1$). The heavy spectrum at tree level has
$M_1 = 3.00$ eV and $M_2 = 3.16$ eV, while three values of
the heaviest mass are considered, $M_3 = 10^{13}\, (10^{14})\, [5\times 10^{14}]$ GeV,
corresponding to the black (dark blue) [light blue] points in the scatter plot.
The vertical red band corresponds to the cosmological constraint, as in Figure~\ref{fig:Ia}.
The dark green contour delimits the region inside which loop-stable points have been found,
while the benchmark of Table~\ref{tab:Ib} is marked in yellow.}
\label{fig:Ib}
\end{figure}
To further explore the parameter space of case Ib, we have
produced numerical seesaw structures qualitatively similar to the benchmark
by following the procedure described while discussing case Ia.
We have specified (at tree level) $M_1 = 3.00$ eV and $M_2 = 3.16$ eV,
and have considered three different values for the heaviest neutrino mass,
$M_3 \in \{10^{13},\, 10^{14},\, 5\times 10^{14}\}$ GeV. 
In Figure~\ref{fig:Ib}
we show the values of the average of $\sin^2 2\vartheta_{\mu e}^{(4)}$ and $\sin^2 2\vartheta_{\mu e}^{(5)}$
against the values of the lightest neutrino mass, for the numerical examples found for case Ib.
The former quantity is expected to represent the order of magnitude of potential signals of this case
in SBL and LBL experiments. Only points for which $\tr\left[m\,m^\dagger\right] \in [0.01,\,1]\,v^2$ are kept.
As before, the dark green contour delimits the region inside which relatively loop-stable points can be found.
Raising the scale of $M_3$ will again lower
the scale of the light neutrino masses.
The approximations used in deriving the oscillation formulae of Eqs.~\eqref{eq:LBL2Ib} and~\eqref{eq:SBL2Ib}
are valid for all the plotted points.

For all numerical examples pertaining to case Ib which are stable under loop corrections,
$BR(\mu \rightarrow e\gamma) \ll 10^{-30}$ is unobservably small,
while one finds $m_\beta < 9.3$ meV and $|m_{\beta\beta}| < 4.6$ meV,
still out of reach of present and near-future experiments.
In the computation of $|m_{\beta\beta}|$, the effects of both eV-scale
neutrinos have been taken into account.
One additionally finds the bounds $|R_{e4}|^2,\,|R_{e5}|^2 \lesssim 0.01$
for the loop-stable numerical examples of this case.

\subsection{Case II: \texorpdfstring{$M_1\ll M_2\sim M_3$}{M1 << M2 \textasciitilde{} M3}}
\begin{table}
\addtocounter{table}{-1}
\renewcommand{\thetable}{\arabic{table}c}
\centering
\renewcommand{\arraystretch}{1.2}
\begin{tabular}{lr}
\toprule
 & {\bf Case II} numerical benchmark \\
\midrule
\addlinespace
$m$ (GeV) &
$\begin{bmatrix*}[r]
-4.15 + 0.47 \,i & ( 4.51 - 1.49 \,i)\times 10^{-9} & (-1.59 + 0.13 \,i)\times 10^{-9} \\
 3.98 + 6.17 \,i & (-5.04 - 4.64 \,i)\times 10^{-9} & ( 1.52 + 2.31 \,i)\times 10^{-9} \\
 1.53 + 6.58 \,i & (-1.90 - 2.68 \,i)\times 10^{-9} & ( 0.59 + 2.59 \,i)\times 10^{-9}
 \end{bmatrix*}$ \\
\addlinespace
$M$ (GeV)  & 
$\begin{bmatrix*}[r]
2.18\times 10^{-6} &  1390               & 2.96               \\
1390               & -2.19\times 10^{-6} & 5.52\times 10^{-7} \\
2.96               &  5.52\times 10^{-7} & 3.33\times 10^{-9}
\end{bmatrix*}$
\\
\addlinespace
\midrule
\addlinespace
$K$  &
$\begin{bmatrix*}[r]
 0.825 +0.061 \,i &  0.536 +0.027 \,i & -0.092 +0.108 \,i \\
-0.302 +0.113 \,i &  0.581 -0.017 \,i &  0.728 -0.052 \,i \\
 0.455 +0.054 \,i & -0.599 +0.075 \,i &  0.651 +0.002 \,i
 \end{bmatrix*}$
\\
\addlinespace
$R$  &
$\begin{bmatrix*}[r]
 0.063 -0.056 \,i & ( 2.11 -0.24 \,i)\times 10^{-3} & (-0.24 -2.11 \,i)\times 10^{-3} \\
-0.066 -0.147 \,i & (-2.03 -3.13 \,i)\times 10^{-3} & (-3.13 +2.03 \,i)\times 10^{-3} \\
-0.021 -0.036 \,i & (-0.79 -3.35 \,i)\times 10^{-3} & (-3.35 +0.79 \,i)\times 10^{-3} 
\end{bmatrix*}$
\\
\addlinespace
$X$  &
$\begin{bmatrix*}[r]
 0.042 +0.014 \,i & 0.007 +0.099 \,i & -0.069 +0.140 \,i \\
( 1.48 +0.64 \,i)\times 10^{-3} & ( 2.30 +0.66 \,i)\times 10^{-4} & (-2.11 +4.89 \,i)\times 10^{-3} \\
(-0.64 +1.48 \,i)\times 10^{-3} & (-0.66 +2.30 \,i)\times 10^{-4} & (-4.89 -2.11 \,i)\times 10^{-3}
\end{bmatrix*}$
\\
\addlinespace
$O_c$ (tree level) \!\!\!\!\!\! \!\!\!\!\!\!\!\!\!\!\!\!& 
$\begin{bmatrix*}[r]
-0.21 +0.62 \,i & -1.02 +0.06 \,i & 0.62 + 0.31 \,i \\
(-1.11 +2.54 \,i)\times 10^4 & (-1.05 +2.42 \,i)\times 10^3 & ( 2.55 +1.12 \,i)\times 10^4 \\
(-2.54 -1.11 \,i)\times 10^4 & (-2.42 -1.05 \,i)\times 10^3 & (-1.12 +2.55 \,i)\times 10^4 
\end{bmatrix*}$
\\
\addlinespace
\midrule
\addlinespace
Masses &
$\begin{matrix*}[l]
m_1 \simeq 4.65\times 10^{-3}\text{ eV}\,,\quad & m_2 \simeq 9.47\times 10^{-3}\text{ eV} \,,\quad & m_3 \simeq 5.01\times 10^{-2}\text{ eV} \,,\,\\ 
M_1 \simeq 1.00\text{ eV}\,,\,               & M_2 \simeq 1390\text{ GeV} \,,\,              & M_3 \simeq 1390\text{ GeV}
\end{matrix*}$ \\
\addlinespace
\midrule
\addlinespace
$3\nu$ $\Delta m^2$ &
$\Delta m^2_\odot = \Delta m^2_{21} \simeq 6.80 \times 10^{-5}\text{ eV}^2\,,\,
\quad\,\,\,\,
\Delta m^2_\text{atm} = \Delta m^2_{31} \simeq 2.48 \times 10^{-3}\text{ eV}^2$
\\
\addlinespace
$3\nu$ mixing angles \!\!\!\!\!\!\!\!\!\!\!\!& 
$\sin^2 \theta_{12} \simeq 0.298\,,\,\quad\,\,\,\,
 \sin^2 \theta_{23} \simeq 0.563\,,\,\quad\,\,\,\,
 \sin^2 \theta_{13} \simeq 0.0212$
\\
\addlinespace
$3\nu$ CPV phases \!\!\!\!\!\! & 
$\delta \simeq 1.32 \pi\,,\,\quad\,\,\,\,
\alpha_{21} \simeq 1.99 \pi\,,\,\quad\,\,\,\,
\alpha_{31} \simeq 0.02 \pi$
\\
\addlinespace
\midrule
\addlinespace
$\sin^2 2 \vartheta^{(i)}_{\mu e}$ & 
$\sin^2 2 \vartheta_{\mu e}^{(4)} \simeq 7.4\times 10^{-4}$
\\
\bottomrule
\end{tabular}
\caption{The same as Table~\ref{tab:Ia} for case II.
For this benchmark, $M_3 - M_2 \simeq 7.6$ eV $\ll M_{2,3}$.}
\label{tab:II}
\end{table}
The numerical data for the benchmark corresponding to this case is given in Table~\ref{tab:II}.
Apart from the three light mostly-active neutrinos,
the spectrum includes a mostly-sterile neutrino with mass $M_1 \sim 1$ eV
and a pair of quasi-degenerate neutrinos with masses $M_2 \simeq M_3 \sim 1$ TeV.
From Table~\ref{tab:II} one sees that the symmetry in the last two columns of $R$ (recall section~\ref{sec:caseII})
is tied to an analogous symmetry in the last two rows of $X$ and of $O_c$.
The latter can be understood from Eqs.~\eqref{eq:epl} and \eqref{eq:xc}.

For the spectrum of case II, one has $n=1$ in Eq.~\eqref{eq:wdef}.
In a LBL context, the expression of Eq.~\eqref{eq:probability}
applied to the transition probability of muon to electron (anti-)neutrinos
can be approximated by:
\begin{equation}
\begin{aligned}
P^\text{LBL}_{\stackon[-.7pt]{$\nu$}{\brabar}_\mu \rightarrow \stackon[-.7pt]{$\nu$}{\brabar}_e}
\,\simeq\,
&\frac{1}{(\Theta\Theta^\dagger)_{\mu\mu}(\Theta\Theta^\dagger)_{ee}}
\Bigg[ 
\left|(\Theta\Theta^\dagger)_{\mu e}\right|^2
-   4 \,\cdot\,\frac{1}{2}\,\re
\left(\Theta_{\mu 4}^*\,\Theta_{e 4}\,\sum_{j=1}^3\,\Theta_{\mu j}\,\Theta_{e j}^*\right)
\\ &
-   4 \sum_{i>j}^3\,\re
\left(\Theta_{\mu i}^*\,\Theta_{e i}\,\Theta_{\mu j}\,\Theta_{e j}^*\right)
\sin^2 \Delta_{ij} 
\pm 2 \sum_{i>j}^{3}\,\im
\left(\Theta_{\mu i}^*\,\Theta_{e i}\,\Theta_{\mu j}\,\Theta_{e j}^*\right)
\sin 2 \Delta_{ij}
\Bigg] \,,
\label{eq:LBL1II}
\end{aligned}
\end{equation}
where terms depending on $\Delta_{4j} \gg 1$ 
have been replaced by their averaged versions.
%
%
If one is in a condition similar to that of
the benchmark of Table~\ref{tab:II},
for which $|(\Theta\Theta^\dagger)_{\mu\mu}(\Theta\Theta^\dagger)_{ee} - 1|$
and $|(\Theta\Theta^\dagger)_{\mu e}|^2$ are negligible,
this expression can be further approximated by:
\begin{align}
P^\text{LBL}_{\stackon[-.7pt]{$\nu$}{\brabar}_\mu \rightarrow \stackon[-.7pt]{$\nu$}{\brabar}_e}
\,\simeq\,
P^{\text{LBL, }3\nu}_{\stackon[-.7pt]{$\nu$}{\brabar}_\mu \rightarrow \stackon[-.7pt]{$\nu$}{\brabar}_e}
+ \frac{1}{2}\sin^2 2 \vartheta^{(4)}_{\mu e}
+ 4\,\re\left(\Theta_{\mu 4}^*\,\Theta_{e 4}\,R_{\mu 2}\,R_{e 2}^*\right)
 \,.
\label{eq:LBL2II}
\end{align}
Here, we have used the unitarity of the full $6\times 6$ mixing matrix,
and the approximate symmetry $R_{\alpha 2}\simeq i\, R_{\alpha 3}$.
If, additionally $|\Theta_{\alpha 4}|^2 = |R_{\alpha 1}|^2  \gg |R_{\alpha 2}|^2 \simeq |R_{\alpha 3}|^2$,
the last term can be neglected and one recovers Eq.~\eqref{eq:LBL2Ia} of case Ia.

%
In a SBL context,
the relevant form of Eq.~\eqref{eq:probability} for
$\stackon[-.7pt]{$\nu$}{\brabar}_\mu \rightarrow \stackon[-.7pt]{$\nu$}{\brabar}_e$
transitions in case II is:
\begin{equation}
\begin{aligned}
P^\text{SBL}_{\stackon[-.7pt]{$\nu$}{\brabar}_\mu \rightarrow \stackon[-.7pt]{$\nu$}{\brabar}_e}
\,\simeq\,
\frac{1}{(\Theta\Theta^\dagger)_{\mu\mu}(\Theta\Theta^\dagger)_{ee}}
\Bigg[ 
\left|(\Theta\Theta^\dagger)_{\mu e}\right|^2
&-   4 \,\re
\left(\Theta_{\mu 4}^*\,\Theta_{e 4}\,\sum_{j=1}^3\,\Theta_{\mu j}\,\Theta_{e j}^*\right)
\sin^2 \Delta_{41} 
\\ &
\pm 2 \,\im
\left(\Theta_{\mu 4}^*\,\Theta_{e 4}\,\sum_{j=1}^{3}\,\Theta_{\mu j}\,\Theta_{e j}^*\right)
\sin 2 \Delta_{41}
\Bigg] \,,
\label{eq:SBL1II}
\end{aligned}
\end{equation}
with $\Delta_{41}\simeq \Delta_{42}\simeq \Delta_{43}$.
One is thus sensitive to oscillations due to the scale of mass-squared differences $\Delta m^2_{4j}$ with $j=1,2,3$,
while the oscillations pertaining to smaller mass-squared differences have not yet developed.
%
%
If one is in a condition similar to that of
the numerical benchmark, this expression can be approximated by:
\begin{align}
P^\text{SBL}_{\stackon[-.7pt]{$\nu$}{\brabar}_\mu \rightarrow \stackon[-.7pt]{$\nu$}{\brabar}_e}
\,\simeq\, 
\left[
\sin^2 2 \vartheta^{(4)}_{\mu e} + 8\,\re\left(\Theta_{\mu 4}^*\,\Theta_{e 4}\,R_{\mu 2}\,R_{e 2}^*\right)
\right]\sin^2 \Delta_{41}
\mp 4\,\im\left(\Theta_{\mu 4}^*\,\Theta_{e 4}\,R_{\mu 2}\,R_{e 2}^*\right)\sin 2 \Delta_{41}
\,,
\label{eq:SBL2II}
\end{align}
where once again the unitarity of the full mixing matrix has been taken into account,
as well as the relation $R_{\alpha 2}\simeq i\, R_{\alpha 3}$. 
If also $|R_{\alpha 1}|^2\gg |R_{\alpha 2}|^2 \simeq |R_{\alpha 3}|^2$,
then the two terms containing $R_{\alpha 2}$ in this equation
can be neglected and one recovers Eq.~\eqref{eq:SBL2Ia} of case Ia.

\begin{figure}[t]
\centering
\includegraphics[width=1.0\linewidth]{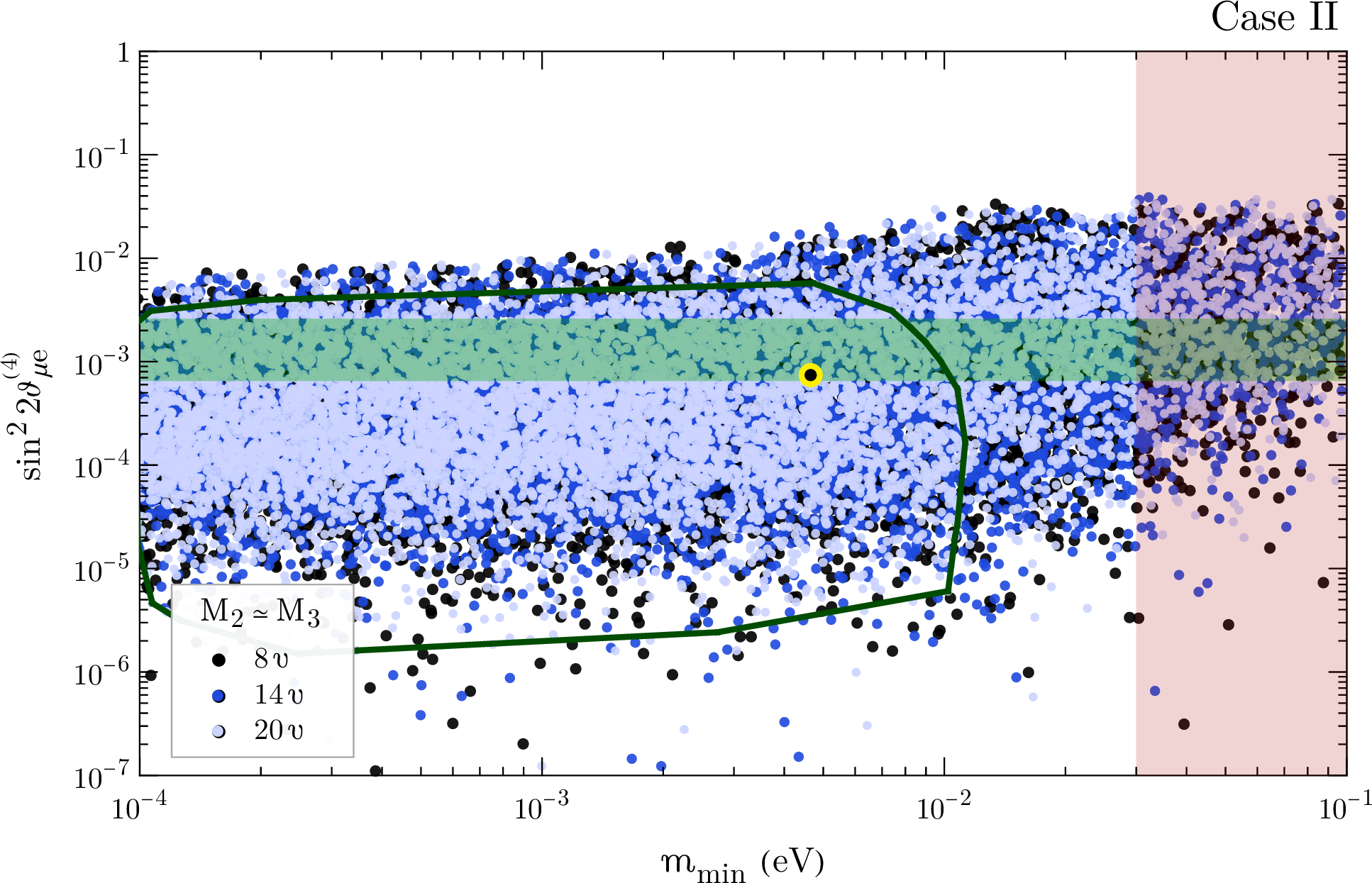}
\caption{
Active-sterile mixing measure $\sin^2 2\vartheta_{\mu e}^{(4)}$ 
{\it versus} the lightest neutrino mass $m_\text{min}$ from a scan of the case-II parameter space,
with NO ($m_\text{min} = m_1$). The heavy spectrum at tree level has
$M_1 = 1$ eV, while three values of the heaviest quasi-degenerate masses are considered,
$M_2 \simeq M_3 = 8\,v\, (14\,v)\, [20\,v]$,
corresponding to the black (dark blue) [light blue] points in the scatter plot.
Here, $v \simeq 174$ GeV is the Higgs VEV.
The horizontal green band shows the $99.7\%$ CL interval of Ref.~\cite{Gariazzo:2015rra},
and the vertical red band corresponds to the cosmological constraint, as in Figure~\ref{fig:Ia}.
The dark green contour delimits the region inside which loop-stable points have been found,
while the benchmark of Table~\ref{tab:II} is marked in yellow.}
\label{fig:II}
\end{figure}
To further explore the parameter space of case II, we have
produced numerical seesaw structures qualitatively similar to our benchmark
by following a procedure similar to that of case Ia.
We have specified (at tree level) $M_1 = 1$ eV
and three different values for the second heaviest neutrino mass,
$M_2\, (\simeq M_3) \in \{8,\,14,\, 20\}\,v$, where $v\simeq 174$ GeV is the Higgs VEV.
We have further scanned the mass splitting $M_3 - M_2$ in the interval $[0.02,\,200]$ eV.
In Figure~\ref{fig:II}
we show the values of $\sin^2 2\vartheta_{\mu e}^{(4)}$ in Eq.~\eqref{eq:sthmue}
against the values of the lightest neutrino mass, for the numerical examples found for case II.
Only points for which $\tr\left[m\,m^\dagger\right] \in [0.001,\,1]\,v^2$ are kept.
As before,
the horizontal green band highlights the range of $\sin^2 2\vartheta_{\mu e}^{(4)}$
preferred by the global fit of Ref.~\cite{Gariazzo:2015rra} and cited at the beginning of the present section,
while the dark green contour delimits the region inside which relatively loop-stable points can be found.
The approximations used in deriving the oscillation formulae of Eqs.~\eqref{eq:LBL2II} and~\eqref{eq:SBL2II}
are valid for all the plotted points.

For the numerical examples pertaining to case II which are stable under loop corrections,
one can obtain values of $BR(\mu \rightarrow e\gamma)$ close to the MEG upper bound of $4.2 \times 10^{-13}$.
Points with larger values of the branching ratio are excluded from our scan.
For the benchmark of Table~\ref{tab:II} one has $BR(\mu \rightarrow e\gamma) \simeq 2.0 \times 10^{-13}$.
Such effects can be probed by the MEG II update~\cite{Cattaneo:2017psr},
which is expected to increase the present sensitivity of MEG by one order of magnitude.
One also finds the bounds $m_\beta < 15$ meV, $|m_{\beta\beta}| < 27$ meV,
and $|R_{e4}|^2 \lesssim 0.02$ for the loop-stable numerical examples of this case.
While KATRIN will seek to improve the current bound on $m_\beta$ down to $0.2$ eV,
values of $|m_{\beta\beta}| \gtrsim 10^{-2}$ eV may be probed in
the next generation of $(\beta\beta)_{0\nu}$-decay experiments~\cite{Vergados:2016hso}.
Concerning the prospect of detecting the heavy neutrino pair in future collider searches,
the reader is further referred to the review~\cite{Antusch:2016ejd}.
If, unlike our benchmark, the heavy neutrino pair would have a mass
in the $1-100$ GeV range and were sufficiently long-lived,
it might lead to displaced vertex signatures~\cite{Antusch:2016vyf}
and produce resolvable neutrino-antineutrino oscillations at colliders~\cite{Antusch:2017ebe}.
Finally, the pseudo-Dirac pair of case II might play a role
in explaining the baryon asymmetry of the Universe
through resonant leptogenesis~\cite{Pilaftsis:2003gt}.
In such a scenario, one should carefully
take into account the washout from the interactions of the
lighter sterile neutrino species.%
\footnote{For an $M_1$ of case II in the range $[0.1,\,50]$ keV,
see the ISS(2,3) analysis of Ref.~\cite{Abada:2017ieq}.}
These interactions may need to be non-standard
in order to reconcile the light sterile neutrino paradigm
with cosmology.

\vskip 2mm
The presented explicit numerical examples are merely illustrative.
However, they give credit to our claim that
models exhibiting an approximate lepton number symmetry 
with at least one sterile neutrino mass at the eV scale
are viable and could play a part in explaining the SBL anomalies.
In the next section we look into CP Violation in the present framework in some detail.

\section{CP Violation in this Framework}
\label{sec:cp}

\subsection{Remarks on CP Violation Measurements}
In order to analyse CP Violation effects,
it is instructive to define CP asymmetries $A_{\nu \overline{\nu}}^{\alpha \beta}$
at the level of oscillation probabilities (see e.g.~\cite{Gandhi:2015xza}):
\begin{align}
A_{\nu \overline{\nu}}^{\alpha \beta}
\,\equiv\,
\frac
{P_{\nu_\alpha \to \nu_\beta} - P_{\overline{\nu}_\alpha \to \overline{\nu}_\beta}}
{P_{\nu_\alpha \to \nu_\beta} + P_{\overline{\nu}_\alpha \to \overline{\nu}_\beta}}
\,\equiv\,
\frac{\Delta P_{\alpha\beta}}
{P_{\nu_\alpha \to \nu_\beta} + P_{\overline{\nu}_\alpha \to \overline{\nu}_\beta}}
\,.
\end{align}
We restrict our discussion to the vacuum case, keeping in mind that in a realistic
context the breaking of CP and CPT due to the asymmetry of the matter which neutrinos traverse
should be taken into account. The requirement of CPT invariance results in the relations
$\Delta P_{\alpha\beta} = -\Delta P_{\beta\alpha}$ and $\Delta P_{\alpha \alpha} = 0$.
From the unitarity of the full mixing matrix, one further has
\begin{align}
\sum_\beta\,\Delta P_{\alpha\beta} = 0\,,
\end{align}
for any $\alpha$, with $\alpha$ and $\beta$ running through the whole index set,
$\alpha,\beta=e$, $\mu$, $\tau$, $s_1,\ldots$, $s_q$.
In a $3\times 3$ unitary context, these relations imply that there is only one
independent difference, which can be chosen as $\Delta P_{e\mu}$.
As shown in~\cite{Gandhi:2015xza}, in a $4\times 4$ unitary framework
they imply the existence of 3 independent differences, say $\Delta P_{e \mu}$,
$\Delta P_{\mu \tau}$, and $\Delta P_{\tau e}$.
In the $6\times 6$ unitary case, we find instead that there are 10 independent differences $\Delta P_{\alpha \beta}$ (see also~\cite{Reyimuaji:2019wbn}), while only the
three of them involving just active neutrinos are experimentally relevant.
Thus, one should generically expect different values for $\Delta P_{e \mu}$,
$\Delta P_{\mu \tau}$, and $\Delta P_{\tau e}$ in a given seesaw-type model.

Using Eq.~\eqref{eq:probability}, with $n$ mostly-sterile neutrinos accessible at 
an oscillation experiment, one finds:
\begin{align}
\Delta P_{\alpha\beta} \,=\,
\frac{4}{(\Theta\Theta^\dagger)_{\alpha\alpha}(\Theta\Theta^\dagger)_{\beta\beta}}
\, \sum_{i>j}^{3+n}\,\im
\left(\Theta_{\alpha i}^*\,\Theta_{\beta i}\,\Theta_{\alpha j}\,\Theta_{\beta j}^*\right)
\sin 2 \Delta_{ij}\,.
\end{align}
Even if none of the new sterile states are accessible -- corresponding to $n=0$ --
one is still expects $\Delta P_{e \mu}$, $\Delta P_{\mu \tau}$, and $\Delta P_{\tau e}$ 
to be independent, as the relevant $3\times 3$ mixing submatrix $\Theta \,(= K)$ is not unitary.
This means that it is possible for CP invariance to hold in one oscillation channel, such as 
$\stackon[-.7pt]{$\nu$}{\brabar}_\mu \rightarrow \stackon[-.7pt]{$\nu$}{\brabar}_e$
and yet be violated in another, such as 
$\stackon[-.7pt]{$\nu$}{\brabar}_\mu \rightarrow \stackon[-.7pt]{$\nu$}{\brabar}_\tau$. 
Indeed, one has:
\begin{align}
\Delta P_{\mu \tau} \,&=\, \Delta P_{e \mu} \,+\,
\frac{4}{\prod_{\alpha = e,\mu,\tau} (\Theta\Theta^\dagger)_{\alpha\alpha}}
\, \sum_{i>j}^{3}\,\im
\Big[\Theta_{\mu i}^*\,\Theta_{\mu j}\Big(
  \Theta_{e i}\,\Theta_{e j}^*      \, (\Theta\Theta^\dagger)_{\tau\tau}
+ \Theta_{\tau i}\,\Theta_{\tau j}^*\, (\Theta\Theta^\dagger)_{ee}
\Big)\Big]
\sin 2 \Delta_{ij}
 \,,\\
\Delta P_{\tau e}   \,&=\, \Delta P_{e \mu} \,-\,
\frac{4}{\prod_{\alpha = e,\mu,\tau} (\Theta\Theta^\dagger)_{\alpha\alpha}}
\, \sum_{i>j}^{3}\,\im
\Big[\Theta_{e i}^*\,\Theta_{e j}\Big(
  \Theta_{\mu i}\,\Theta_{\mu j}^*  \, (\Theta\Theta^\dagger)_{\tau\tau}
+ \Theta_{\tau i}\,\Theta_{\tau j}^*\, (\Theta\Theta^\dagger)_{\mu\mu}
\Big)\Big]
\sin 2 \Delta_{ij}
 \,.
\end{align}
It is then possible to have a zero $\Delta P_{e\mu}$ while $\Delta P_{\mu \tau }$ and/or $\Delta P_{\tau e}$ are non-zero.
Notice that if $\Theta$ here were unitary, one would recover $\Delta P_{e \mu} = \Delta P_{\mu \tau} = \Delta P_{\tau e}$.
Thus, deviations from unitarity are a potential source of CP Violation. This should come as no surprise,
if one recalls that $\eta$ in Eq.~\eqref{eq:khu} is a complex hermitian matrix containing, in general, CPV physical phases.

For the cases analysed in sections~\ref{sec:loopandsym} and~\ref{sec:numeric}, 
one has $n = 1,2$. Explicit expressions for the CP asymmetries relevant in a SBL context can be obtained
from the approximate relations~\eqref{eq:SBL2Ib} and~\eqref{eq:SBL2II} of cases Ib and II, respectively.
Instead, from the relation \eqref{eq:SBL2Ia} one sees that SBL CP asymmetries for case Ia are negligible.
One has, for case Ib:
\begin{align}
\Delta P_{e\mu}^\text{SBL,\:Ib} \,\simeq\,
4 \,\im\left(\Theta_{\mu 4}^*\,\Theta_{e 4}\,\Theta_{\mu 5}\,\Theta_{e 5}^*\right)\sin 2 \Delta_{54}
\,,
\end{align}
while for case II:
\begin{align}
\Delta P_{e\mu}^\text{SBL,\:II} \,\simeq\,
8 \,\im\left(\Theta_{\mu 4}^*\,\Theta_{e 4}\,R_{\mu 2}\,R_{e 2}^*\right)\sin 2 \Delta_{41}
\,.
\end{align}

\subsection{CP-odd Weak Basis Invariants}
\label{sec:WBinv}

In section \ref{sec:param} we have shown (see also Ref.~\cite{Branco:2001pq})
that in the present framework, where three right-handed neutrinos have been added to the SM,
there are 6 CPV phases. They can be made to appear in the Dirac mass matrix $m$
by changing to the weak basis (WB) where the charged lepton mass matrix $m_l$
and the Majorana mass matrix $M$ are diagonal and real.
In the study of CP Violation, it is very useful to construct CP-odd WB invariants
following the procedure introduced for the first time for the quark sector
in Ref.~\cite{Bernabeu:1986fc}, see also~\cite{Branco:1999fs}.
This procedure was later applied by different
authors~\cite{Branco:1986gr,Pilaftsis:1997jf,Branco:1998bw,Branco:2001pq,%
Davidson:2003yk,Branco:2005jr,Dreiner:2007yz,Wang:2014lla}
to the leptonic sector, in order to build CP-odd WB invariants relevant in
several different contexts.
Such invariants can be calculated in any convenient WB and their non-vanishing
signals the presence of CP-breaking.
We define six WB invariants which are sensitive to the leptonic CPV phases:%
\begin{equation}
\begin{aligned}
i_R \,&=\, \im \,\tr \left[ M^\dagger\, M \, m^\dagger \,m\, (M^\dagger\, M)^2\, (m^\dagger\, m)^2 \right]\,, \\[2mm]
j_R^{(1)} \,&=\, \im \,\tr \left[ M^{-1}\,m^T\,m^*\,M\,m^\dagger\,m \right]\,, \\[2mm]
j_L^{(1)} \,&=\, \im \,\tr \left[ M^\dagger\,M\,m^\dagger\,h_\ell\,m\,m^\dagger\,m \right]\,,
\end{aligned}
\qquad
\begin{aligned}
i_L \,&=\, \im \tr \,\left[ h_\ell \, m \,m^\dagger\, h_\ell^{2}\, (m\, m^\dagger)^2\right]\,,\\[2mm]
j_R^{(2)} \,&=\, \im \,\tr \left[ M^{-1}\,m^T\,m^*\,M\,(m^\dagger\,m)^2\right]\,,\\[2mm]
j_L^{(2)} \,&=\, \im \,\tr \left[ M^\dagger\,M\,m^\dagger\,h_\ell\,m\,(m^\dagger\,m )^2 \right]\,,
\end{aligned}
\label{eq:js}
\end{equation}
where we have assumed $M$ to be invertible and have additionally defined $h_\ell \equiv m_l\,m_l^\dagger$.

To see how the above invariants capture the 6 leptonic CPV phases, consider the
aforementioned WB of $m_l$ and $M$ diagonal and real:
$h_\ell = \diag(m_e^2,\, m_\mu^2,\, m_\tau^2)$ and
$M = \tilde{D} = \diag(\tilde{M}_1,\, \tilde{M}_2,\, \tilde{M}_3)$.
Recall that in this basis the full neutrino mass matrix $\mathcal{M}$ is not diagonal
and therefore the $\tilde{M}_i$ do not coincide with the physical masses $M_i$.
We further consider the singular value decomposition of $m$: 
\begin{align}
m\,=\,V_L\,d_m\,V_R\,,
\end{align}
with $V_{L,R}$ unitary and $d_m = \diag({d}_1,\,{d}_2,\,{d}_3)$ real and positive.
The 6 physical CPV phases of interest are contained in $m$, since 3 out of
its original 9 can be removed by rephasing left-handed fields.
A parametrisation of $V_L$ and $V_R$ which captures explicitly these phases is: 
\begin{align}
V_L\,=\,V_{\delta_L}\,K_L\,,
\qquad
V_R\,=\,V_{\delta_R}\,K_R\,,
\end{align}
with $K_{L,R}\equiv \diag(1,\, e^{i\alpha_{L,R}},\, e^{i\beta _{L,R}})$ and 
\begin{align}
V_{\delta_{L,R}}\,\equiv \,O_{23}\,\diag(1,1,e^{i\delta
_{L,R}})\,O_{13}\,O_{12}\,,
\end{align}
the $O_{ij}$ being ordinary real rotation matrices in the $i$-$j$ plane, e.g.
\begin{align}
O_{23}(\theta_{23L})\,=\,
\begin{pmatrix}
1 & 0 & 0 \\ 
0 &  \cos \theta_{23L} & \sin \theta_{23L} \\ 
0 & -\sin \theta_{23L} & \cos \theta_{23L}
\end{pmatrix}\,.
\end{align}
The phases of interest are then manifestly $\alpha_{L,R}$, $\beta_{L,R}$
and $\delta_{L,R}$.
Using this parametrisation, the invariants can be cast in the forms:
\begin{equation}
\begin{aligned}
i_R       \,&=\, \mathcal{K}_{i_R}\, \sin \delta_R \,, \quad
i_L       \, =\, \mathcal{K}_{i_L}\, \sin \delta_L \,,\\[3mm]
j_R^{(a)} \,&=\, \mathcal{K}_{j_R^{(a)}}^{\delta_R}\, \sin \delta_R 
           \,+\, \mathcal{K}_{j_R^{(a)}}^{2\alpha_R}\, \sin 2\alpha_R
           \,+\, \mathcal{K}_{j_R^{(a)}}^{2\beta_R} \, \sin 2\beta_R \,, \\[3mm]
j_L^{(a)} \,&=\, \mathcal{K}_{j_L^{(a)}}^{\delta_R}\, \sin \delta_R 
           \,+\, \mathcal{K}_{j_L^{(a)}}^{\delta_L}\, \sin \delta_L 
           \,+\, \mathcal{K}_{j_L^{(a)}}^{\alpha_L}\, \sin \alpha_L
           \,+\, \mathcal{K}_{j_L^{(a)}}^{\beta_L} \, \sin \beta_L \,,
\label{eq:js1}
\end{aligned}
\end{equation}
with $a=1,2$. Explicit expressions for the $\mathcal{K}$ coefficients
are given in Appendix~\ref{app:WBinv}.
It is clear that $i_R$ and $i_L$ are sensitive to CPV values of $\delta_R$ and $\delta_L$, respectively,
while the $j_R^{(1,2)}$ ($j_L^{(1,2)}$) are further sensitive to
$\alpha_R$ and $\beta_R$ ($\alpha_L$ and $\beta_L$).

\section{Summary and Conclusions}
\label{sec:conclusions}

We have seen that in the framework of the type-I seesaw mechanism
one can naturally have at least one sterile neutrino with a mass
of around one eV.
This can be inferred using a general exact parametrisation,
defined in \cite{Agostinho:2017wfs}, that is valid irrespectively
of the size and structure of the neutrino mass matrix.
Thus we are able to analyse a general seesaw where not all
of the three mostly\discretionary{-}{-}{-}sterile neutrinos need to be very
heavy. We have focused on models where at least one of the
sterile neutrinos is light and its mixing with the active
neutrinos is small enough to respect experimental bounds but
sufficiently large to be relevant to low energy phenomenology
-- for instance, providing a natural explanation to the
short-baseline anomalies.

In section~\ref{sec:framework}, we have shown how the usual 
seesaw formulae have to be generalised in order to be applicable
to the special region of parameters which we are considering.
In particular, we have written the full neutrino mixing matrix
in terms of a $3\times 3$ unitary matrix and a $3\times 3$ 
general complex matrix, which encodes the deviations from unitarity.
The latter was further parametrised at tree level in terms of
neutrino masses and a complex orthogonal matrix.
We carefully distinguish approximate and exact relations,
which are valid in any seesaw regime. 
Namely, we have found an exact formula for the neutrino
Dirac mass matrix $m$ in terms of neutrino masses, neutrino mixing
and deviations from unitarity, which generalises the usual
Casas-Ibarra parametrisation of $m$. We additionally derive 
an exact seesaw-like relation, equating the product of neutrino
masses and the square of the absolute value of $\det\, m$.

In section~\ref{sec:devunit}, we have further discussed
the parametrisation of deviations from unitarity
as well as constraints on said deviations in our framework.
These significantly depend on the masses of the heavy neutrinos.
In this context, we also find a bound on the lightest neutrino mass $m_\text{min}$,
useful whenever a light sterile is present in the seesaw spectrum.
For the cases of interest, with an eV-scale sterile neutrino and large deviations
from unitarity, one has $m_\text{min} \lesssim 0.1$ eV.

In sections~\ref{sec:loopandsym} and~\ref{sec:numeric} we give examples of viable textures
with at least one sterile neutrino with a mass at the eV scale.
Such light sterile states arise naturally by imposing an approximately conserved lepton number symmetry.
Before the breaking, and for an appropriate assignment of leptonic charges, the lightest
neutrinos are massless at tree level. After the breaking, the lightest neutrinos acquire calculable masses,
with mass differences in agreement with experiment, after the relevant one-loop correction to the zero block
of the neutrino mass matrix has been taken into account. This correction is cast in a simple form,
highlighting the cancellations required by radiative stability, in section~\ref{sec:loop}.
We identify two symmetric textures (I and II) of the neutrino mass matrix which
allow for a separation of high (TeV -- GUT) and low ($\lesssim$ keV) scales. 
We then concentrate on three particular scenarios, with differing spectra $(M_1,\, M_2,\,M_3)$ of heavy neutrinos:
case Ia, for which $M_1\ll M_2\ll M_3$; case Ib, with $M_1\sim M_2\ll M_3$; and case II, where $M_1\ll M_2\sim M_3$.
Numerical benchmarks are given for each of these three cases in Tables~\ref{tab:Ia}\,--\,\ref{tab:II}.
Related regions in parameter space are explored in Figures~\ref{fig:Ia}\,--\,\ref{fig:II},
which show that these models can accommodate enough active-sterile mixing
to play a role in the explanation of short\discretionary{-}{-}{-}baseline anomalies.
Since the formulae for neutrino oscillation probabilities are modified
in the presence of deviations from unitarity,
we present, for each case, approximate expressions for muon to electron (anti-)neutrino transition probabilities,
quantifying the impact of light sterile states on oscillations, for both short- and long-baseline experiments.
Attention is further given to the future testability of the proposed models
through non-oscillation effects of the extra neutrino states.

We conclude our work in section~\ref{sec:cp} by discussing CP Violation in the type-I seesaw framework
under analysis. 
At the level of oscillation probability asymmetries,
we have found that deviations from unitarity may source CP Violation,
with generically independent effects in the standard transition channels.
We have also constructed 6 CP-odd weak basis invariants which are sensitive the CP-violating phases
in the lepton sector. This last point has been shown explicitly,
for a particular choice of weak basis and parametrisation of $m$.

\section*{Acknowledgements}
This work was supported by Funda\c{c}\~{a}o para a Ci\^{e}ncia e a Tecnologia (FCT, Portugal)
through the projects CFTP-FCT Unit 777 (UID/FIS/00777/2013 and UID/FIS/00777/2019), 
CERN/FIS-PAR/0004/2017, and PTDC/FIS-PAR/29436/2017 which are partially funded
through POCTI (FEDER), COMPETE, QREN and EU. The work of PMFP was supported by
several short term Master-type fellowships from the CFTP and CERN projects listed above. At
present PMFP acknowledges support from FCT through PhD grant SFRH/BD/145399/2019.
The work of JTP has been supported by the PTDC project listed above. GCB and MNR thank
the CERN Theoretical Physics Department, where part of this work was done, for hospitality
and partial support and also acknowledge participation in the CERN Neutrino Platform --
Theory working group (CENF-TH).

\appendix

\section{Explicit Expressions for Weak Basis Invariants}
\label{app:WBinv}
{
\def\at{\alpha_R}
\def\bt{\beta_R}
\def\dt{\delta_R}
\def\ao{\alpha_L}
\def\bo{\beta_L}
\def\do{\delta_L}
\def\tm{\tilde M}
\def\th{\theta}
\def\dm{{d}}
\def\hle{m_e^2}
\def\hlm{m_\mu^2}
\def\hlt{m_\tau^2}

Using the definitions of section~\ref{sec:WBinv}, and in the WB there considered,
the WB invariants of Eq.~\eqref{eq:js} read:
\begin{equation}
\begin{aligned}
i_R &= 
 \frac{1}{2i}\,
 \sum_{ijkl}\,
 \tilde M_i^2 \,\tilde M_k^4 \,d_j^2\, d_l^2\, (d_l^2 - d_j^2)
\,\left(V_{\delta_R}\right)_{jk}
\,\left(V_{\delta_R}\right)_{li}
\,\left(V_{\delta_R}\right)^*_{ji}
\,\left(V_{\delta_R}\right)^*_{lk}
\\
&=
 \frac{1}{2i}\,
 \sum_{ijkl}\,
 \tilde M_i^2 \,\tilde M_k^2 \,(\tilde M_k^2 - \tilde M_i^2) \,d_j^2\, d_l^4
\,\left(V_{\delta_R}\right)_{jk}
\,\left(V_{\delta_R}\right)_{li}
\,\left(V_{\delta_R}\right)^*_{ji}
\,\left(V_{\delta_R}\right)^*_{lk}\,,
\end{aligned}
\end{equation}
\begin{equation}
\begin{aligned}
i_L &=  
 \frac{1}{2i}\,
 \sum_{ijkl}\,
 (h_\ell)_l \,(h_\ell)_j^2 \,d_i^2\, d_k^2\, (d_k^2 - d_i^2)
\,\left(V_{\delta_L}\right)_{jk}
\,\left(V_{\delta_L}\right)_{li}
\,\left(V_{\delta_L}\right)^*_{ji}
\,\left(V_{\delta_L}\right)^*_{lk}
\\
&=
 \frac{1}{2i}\,
 \sum_{ijkl}\,
  (h_\ell)_j \, (h_\ell)_l \,\left[ (h_\ell)_j -  (h_\ell)_l\right] \,d_i^2\, d_k^4
\,\left(V_{\delta_L}\right)_{jk}
\,\left(V_{\delta_L}\right)_{li}
\,\left(V_{\delta_L}\right)^*_{ji}
\,\left(V_{\delta_L}\right)^*_{lk}\,,\\
\end{aligned}
\end{equation}
\begin{align}
j_R^{(1)} &= 
\frac{1}{2i}\,\sum_{ijkl}\,
 d^{2}_i\, d_k^2\,
 \frac{\tilde M_l^2-\tilde M_j^2}{\tilde M_l\tilde M_j}\,
 (K_{R})_j^2\,(K_{R}^*)^{2}_l
\,\left(V_{\delta_R}\right)_{ij}
\,\left(V_{\delta_R}\right)_{kj}
\,\left(V_{\delta_R}\right)^*_{il}
\,\left(V_{\delta_R}\right)^*_{kl}\,,\\
j_R^{(2)} &= 
\frac{1}{2i}\,\sum_{ijkl}\,
 d^{4}_i \, d_k^2\,
 \frac{\tilde M_l^2-\tilde M_j^2}{\tilde M_l\tilde M_j}\,
 (K_{R})_j^2\,(K_{R}^*)^{2}_l
\,\left(V_{\delta_R}\right)_{ij}
\,\left(V_{\delta_R}\right)_{kj}
\,\left(V_{\delta_R}\right)^*_{il}
\,\left(V_{\delta_R}\right)^*_{kl}\,,
\end{align}
\begin{align}
j_L^{(1)} &= 
 \frac{1}{2i}\,\sum_{ijkl}\,d_i \,d_k \,\left(d^{2}_i-d^{2}_k\right)\,
 \tilde M_j^2 \,(h_\ell)_l\,
 (K_{L})_i\,(K_{L}^*)_k
\,\left(V_{\delta_R}\right)_{ij}
\,\left(V_{\delta_R}\right)^*_{kj}
\,\left(V_{\delta_L}\right)_{li}
\,\left(V_{\delta_L}\right)^*_{lk}\,,\\
j_L^{(2)} &= 
 \frac{1}{2i}\,\sum_{ijkl}\,d_i \,d_k \,\left(d^{4}_i-d^{4}_k\right)\,
 \tilde M_j^2 \,(h_\ell)_l\,
 (K_{L})_i\,(K_{L}^*)_k
\,\left(V_{\delta_R}\right)_{ij}
\,\left(V_{\delta_R}\right)^*_{kj}
\,\left(V_{\delta_L}\right)_{li}
\,\left(V_{\delta_L}\right)^*_{lk}\,.
\end{align}

Using further the given parametrisations of $V_{\delta_{L,R}}$ and $K_{L,R}$,
one obtains the result of Eq.~\eqref{eq:js1}, with:
\begin{align}
&\begin{aligned}
\mathcal{K}_{i_R} \,&=\, 
-\frac{1}{8} 
\cos \theta_{13R}
\, \sin 2 \theta_{12R}
\, \sin 2 \theta_{13R}
\, \sin 2 \theta_{23R}
\\ & \quad \times
\left(\dm_1^2-\dm_2^2\right)
\left(\dm_2^2-\dm_3^2\right)
\left(\dm_3^2-\dm_1^2\right)
\left(\tm_1^2-\tm_2^2\right)
\left(\tm_2^2-\tm_3^2\right)
\left(\tm_3^2-\tm_1^2\right)\,,
\end{aligned}\\[3mm]&
\begin{aligned}
\mathcal{K}_{i_L} \,&=\, 
\frac{1}{8} 
\cos \theta_{13L}
\, \sin 2 \theta_{12L}
\, \sin 2 \theta_{13L}
\, \sin 2 \theta_{23L}
\\ &\quad \times
\left(\dm_1^2-\dm_2^2\right)
\left(\dm_2^2-\dm_3^2\right)
\left(\dm_3^2-\dm_1^2\right)
\left(\hle-\hlm\right)
\left(\hlm-\hlt\right)
\left(\hlt-\hle\right)\,,
\end{aligned}\\[3mm]&
\begin{aligned}
\mathcal{K}_{j_R^{(a)}}^{\delta_R} \,&=\, 
\Delta^{(a)}_2\, \cos 2 \at\, \cos 2 \bt
+2\,\Delta^{(a)}_3\, \cos 2 \at \, \cos 2 \bt \, \cos \dt
\\&\,\,
- \Delta^{(a)}_6 \,\frac{8 \cos 2 \theta_{12R}}{\cos 4 \theta_{12R}+3} \,\cos 2 \at\, \cos \dt
- \Delta^{(a)}_7 \,\frac{\cos 2 \at}{\cos 2 \theta_{12R}}
\\&\,\,
+\Delta^{(a)}_8\, \cos 2 \bt
+2\,\Delta^{(a)}_9\, \cos 2 \bt \, \cos \dt
\,,
\end{aligned}\\[3mm]&
\begin{aligned}
\mathcal{K}_{j_R^{(a)}}^{2 \alpha_R} \,&=\, 
- \Delta^{(a)}_2\, \cos\,(2\bt+\dt)
- \Delta^{(a)}_3\, \cos 2 (\bt + \dt)
+ \Delta^{(a)}_4\, \cos 2 \bt
+ \Delta^{(a)}_5 
\\&\,\,
+ \Delta^{(a)}_6 \cos \dt + \Delta^{(a)}_7 \cos 2 \dt 
\,,
\end{aligned}\\[3mm]&
\begin{aligned}
\mathcal{K}_{j_R^{(a)}}^{2 \beta_R} \,&=\,
\Delta^{(a)}_1
+ \Delta^{(a)}_2\, \cos 2 \at \, \cos \dt
+ \Delta^{(a)}_3\, \cos 2 \at \, \cos 2 \dt
- \Delta^{(a)}_4\, \cos 2 \at
\\&\,\,
+ \Delta^{(a)}_8\, \cos \dt
+ \Delta^{(a)}_9\, \cos 2 \dt
\,,
\end{aligned}\\[3mm]&
\begin{aligned}
\mathcal{K}_{j_L^{(a)}}^{\delta_R} \,&=\,
  \Delta_{3}^{\prime\, (a)} \, \cos \,(\ao-\bo)
+ \Delta_{4}^{\prime\, (a)} \, \cos 2 \th_{12L} \, \cos \do \,  \cos \ao 
- \Delta_{5}^{\prime\, (a)} \, \cos \bo \, \cos \do
\\&\,\,
+ \Delta_{9}^{\prime\, (a)} \, \cos \ao
- \Delta_{10}^{\prime\, (a)} \, \frac{ \cos \,(\ao-\bo-\do) }{\cos 2 \th_{23R}}
+ \Delta_{12}^{\prime\, (a)} \, \cos \bo
\,,
\end{aligned}	\\[3mm]&
\begin{aligned}
\mathcal{K}_{j_L^{(a)}}^{\delta_L} \,&=\, 
 \Delta_{2}^{\prime\, (a)} \, \cos \ao \, \cos \bo
-\Delta_{4}^{\prime\, (a)} \, \cos \, (\ao+\dt) 
-\Delta_{5}^{\prime\, (a)} \, \cos \bo \, \cos \dt
 \\&\,\,
- \Delta_{7}^{\prime\, (a)} \, \frac{\cos \ao}{\cos 2 \th_{12L}} 
- \Delta_{10}^{\prime\, (a)} \, \cos \dt \, \cos \ao \, \cos \bo
+ \Delta_{11}^{\prime\, (a)} \, \cos \bo
\,,
\end{aligned}\\[3mm]&
\begin{aligned}
\mathcal{K}_{j_L^{(a)}}^{\alpha_L} \,&=\,
- \Delta_{2}^{\prime\, (a)} \, \cos \,(\bo+\do)
- \Delta_{3}^{\prime\, (a)} \, \cos 2 \th_{23R}\, \cos \dt\, \cos \bo
 \\&\,\,
+ \Delta_{4}^{\prime\, (a)} \, \cos 2 \th_{12L} \, \cos \do \, \cos \dt 
+ \Delta_{6}^{\prime\, (a)} \, \cos \bo
+ \Delta_{7}^{\prime\, (a)} \cos \do
+ \Delta_{8}^{\prime\, (a)}
 \\&\,\,
+ \Delta_{9}^{\prime\, (a)} \, \cos \dt
+ \Delta_{10}^{\prime\, (a)} \, \cos \dt \, \cos \,(\bo+\do)
\,,
\end{aligned}\\[3mm]&
\begin{aligned}
\mathcal{K}_{j_L^{(a)}}^{\beta_L} \,&=\,
 \Delta_{1}^{\prime\, (a)} \, \sin \bo
+\Delta_{2}^{\prime\, (a)} \, \cos \ao \, \cos \do
+\Delta_{3}^{\prime\, (a)} \, \cos 2 \th_{23R}\, \cos \dt\, \cos \ao
 \\&\,\,
-\Delta_{5}^{\prime\, (a)} \, \cos \,(\do+\dt)
-\Delta_{6}^{\prime\, (a)} \, \cos \ao
- \Delta_{10}^{\prime\, (a)} \, \cos \dt \, \cos \ao \, \cos \do
 \\&\,\,
+ \Delta_{11}^{\prime\, (a)} \, \cos \do
+ \Delta_{12}^{\prime\, (a)} \, \cos \dt
\,.
\end{aligned}
\end{align}
It should be noted that the $\Delta_i^{(\prime)(a)}$, where $a=1,2$,
are independent of the 6 leptonic CPV phases $\delta_{R,L}$, $\alpha_{R,L}$, and $\beta_{R,L}$.
The rather lengthy expressions for these quantities are finally given below:
\begin{align*}
%
%
\Delta^{(1)}_1 &=
\frac{1}{16}\,
\frac{\tm_1^2 - \tm_3^2}{\tm_1 \tm_3}\,
\cos^2 \th_{12R}\, \sin^2 \th_{13R}
\left[
 \left(\dm_1^2-\dm_2^2\right)
+\left(\dm_1^2-\dm_3^2\right)
+ \cos 2 \th_{23R}\left(\dm_2^2-\dm_3^2\right)
\right]^2\,,\\
%
%
\Delta^{(1)}_2 &=
\frac{1}{8}\,
\frac{\tm_2^2 - \tm_3^2}{\tm_2 \tm_3}\,
\cos \th_{13R}\, \sin 2\th_{13R}\,\sin 2\th_{23R}\,\sin 2\th_{12R}
\left(\dm_2^2-\dm_3^2\right)
\nonumber \\ &\qquad \times
\left[
 \left(\dm_1^2-\dm_2^2\right)
+\left(\dm_1^2-\dm_3^2\right)
+ \cos 2 \th_{23R}\left(\dm_2^2-\dm_3^2\right)
\right]\,,\\
%
%
\Delta^{(1)}_3 &=
\frac{1}{4}\,
\frac{\tm_2^2 - \tm_3^2}{\tm_2 \tm_3}\,
\cos^2 \th_{13R}\, \cos^2 \th_{12R}\,\sin^2 2\th_{23R}
\left(\dm_2^2-\dm_3^2\right)^2\,,\\
%
%
\Delta^{(1)}_4 &=
-\frac{1}{16}\,
\frac{\tm_2^2 - \tm_3^2}{\tm_2 \tm_3}\,
\sin^2 \th_{12R}\, \sin^2 2\th_{13R}
\left[
 \left(\dm_1^2-\dm_2^2\right)
+\left(\dm_1^2-\dm_3^2\right)
+ \cos 2 \th_{23R}\left(\dm_2^2-\dm_3^2\right)
\right]^2\,,\\
%
%
\Delta^{(1)}_5 &=
\frac{1}{128}\,
\frac{\tm_1^2 - \tm_2^2}{\tm_1 \tm_2}\,
\sin^2 2\th_{12R}
\nonumber \\ &\qquad \times
\left[
12 \left(\dm_1^2-\dm_2^2\right)\left(\dm_1^2-\dm_3^2\right)
- 5\left(\dm_2^2-\dm_3^2\right)^2
\right.
\nonumber \\ &\qquad\quad\, 
\left.
+\cos 2 \th_{23R} \left(\dm_2^2-\dm_3^2\right)  \left[27 \cos 2 \th_{23R}  \left(\dm_2^2-\dm_3^2\right) - 10 \left(\left(\dm_1^2-\dm_2^2\right)
+\left(\dm_1^2-\dm_3^2\right)\right)\right]
\right.
\nonumber \\ &\qquad\quad\, 
\left.
+
2 \cos 2 \th_{13R}
\left(
\left(\dm_2^2-\dm_3^2\right)^2+8 \left(\dm_1^2-\dm_2^2\right) \left(\dm_1^2-\dm_3^2\right)
-5 \cos 4 \th_{23R} \left(\dm_2^2-\dm_3^2\right)^2 
\right.\right.
\nonumber \\ &\qquad\qquad\qquad\qquad\qquad\qquad\qquad\quad\,\,\,
-4 \cos 2 \th_{23R} \left(\left(\dm_1^2-\dm_2^2\right)+\left(\dm_1^2-\dm_3^2\right)\right) \left(\dm_2^2-\dm_3^2\right) 
\Big)
\nonumber \\ &\qquad\quad\, 
+ \cos 4 \th_{13R}
\left( \left(\dm_1^2-\dm_2^2\right)
+\left(\dm_1^2-\dm_3^2\right)
+ \cos 2 \th_{23R}\left(\dm_2^2-\dm_3^2\right)
\right)^2
\Big]\,,\\
%
%
\Delta^{(1)}_6 &=
\frac{1}{16}\,
\frac{\tm_1^2 - \tm_2^2}{\tm_1 \tm_2}\,
\sin 4 \th_{12R}\, \sin \th_{13R}\,\sin 2 \th_{23R}
\left(\dm_2^2-\dm_3^2\right)
\nonumber \\ &\qquad \times
\left[
(3-\cos 2 \th_{13R})\cos 2 \th_{23R} \left(\dm_2^2-\dm_3^2\right) 
-2\cos^2 \th_{13R} \left(\left( \dm_1^2-\dm_2^2\right)+\left(\dm_1^2-\dm_3^2\right) \right) 
\right]\,,\\
%
%
\Delta^{(1)}_7 &=
\frac{1}{16}\,
\frac{\tm_1^2 - \tm_2^2}{\tm_1 \tm_2}\,
\left(3+\cos 4 \th_{12R}\right)\,\sin^2 \th_{13R}\,\sin^2 2\th_{23R}
\left(\dm_2^2-\dm_3^2\right)^2\,,\\
%
%
\Delta^{(1)}_8 &=
-\frac{1}{8}\,
\frac{\tm_1^2 - \tm_3^2}{\tm_1 \tm_3}\,
\cos \th_{13R}\, \sin 2\th_{13R}\,\sin 2\th_{23R}\,\sin 2\th_{12R}
\left(\dm_2^2-\dm_3^2\right)
\nonumber \\ &\qquad \times
\left[
 \left(\dm_1^2-\dm_2^2\right)
+\left(\dm_1^2-\dm_3^2\right)
+ \cos 2 \th_{23R}\left(\dm_2^2-\dm_3^2\right)
\right]\,,\\
%
%
\Delta^{(1)}_9 &=
\frac{1}{4}\,
\frac{\tm_1^2 - \tm_3^2}{\tm_1 \tm_3}\,
\cos^2 \th_{13R}\,\sin^2 \th_{12R}\,\sin^2 2\th_{23R}
\left(\dm_2^2-\dm_3^2\right)^2\,.
\end{align*}

\begin{align*}
%
%
\Delta^{(2)}_1 &=
\frac{1}{16}\,
\frac{\tm_1^2 - \tm_3^2}{\tm_1 \tm_3}\,
\cos^2 \th_{12R}\, \sin^2 \th_{13R} 
\nonumber \\ &\qquad \times
\left[
 \left(\dm_1^2-\dm_2^2\right)
+\left(\dm_1^2-\dm_3^2\right)
+ \cos 2 \th_{23R}\left(\dm_2^2-\dm_3^2\right)
\right]
\nonumber \\ &\qquad \times
\left[
 \left(\dm_1^4-\dm_2^4\right)
+\left(\dm_1^4-\dm_3^4\right)
+ \cos 2 \th_{23R}\left(\dm_2^4-\dm_3^4\right)
\right]\,,\\
%
%
\Delta^{(2)}_2 &=
\frac{1}{8}\,
\frac{\tm_2^2 - \tm_3^2}{\tm_2 \tm_3}\,
\cos \th_{13R}\, \sin 2\th_{13R}\,\sin 2\th_{23R}\,\sin 2\th_{12R}
\left(\dm_2^2-\dm_3^2\right)
\nonumber \\ &\qquad \times
\left[
 \left(\dm_1^4-\dm_2^4\right)
+\left(\dm_1^4-\dm_3^4\right)
-\left(\dm_1^2-\dm_2^2\right)\left(\dm_1^2-\dm_3^2\right)
+ \cos 2 \th_{23R}\left(\dm_2^4-\dm_3^4\right)
\right]\,,\\
%
%
\Delta^{(2)}_3 &=
\frac{1}{4}\,
\frac{\tm_2^2 - \tm_3^2}{\tm_2 \tm_3}\,
\cos^2 \th_{13R}\, \cos^2 \th_{12R}\,\sin^2 2\th_{23R}
\left(\dm_2^2-\dm_3^2\right)\left(\dm_2^4-\dm_3^4\right)\,,\\
%
%
\Delta^{(2)}_4 &=
-\frac{1}{16}\,
\frac{\tm_2^2 - \tm_3^2}{\tm_2 \tm_3}\,
\sin^2 \th_{12R}\, \sin^2 2\th_{13R}
\nonumber \\ &\qquad \times
\left[
 \left(\dm_1^2-\dm_2^2\right)
+\left(\dm_1^2-\dm_3^2\right)
+ \cos 2 \th_{23R}\left(\dm_2^2-\dm_3^2\right)
\right]
\nonumber \\ &\qquad \times
\left[
 \left(\dm_1^4-\dm_2^4\right)
+\left(\dm_1^4-\dm_3^4\right)
+ \cos 2 \th_{23R}\left(\dm_2^4-\dm_3^4\right)
\right]\,,\\
%
%
\Delta^{(2)}_5 &=
\frac{1}{256}\,
\frac{\tm_1^2 - \tm_2^2}{\tm_1 \tm_2}\,
\sin^2 2\th_{12R}
\nonumber \\ &\qquad \times
\left[
12 \left(\left(\dm_1^2-\dm_2^2\right) \left(\dm_1^4-\dm_2^4\right)+\left(\dm_1^2-\dm_3^2\right) \left(\dm_1^4-\dm_3^4\right)\right)+5 \left(\dm_2^2-\dm_3^2\right) \left(\dm_2^4-\dm_3^4\right)
\right.
\nonumber \\ &\qquad\quad\, 
\left.
+27 \cos 4 \th_{23R} \left(\dm_2^2-\dm_3^2\right) \left(\dm_2^4-\dm_3^4\right) 
\right.
\nonumber \\ &\qquad\quad\, 
\left.
+20  \cos 2 \th_{23R} \left(\left(\dm_1^2-\dm_2^2\right) \left(\dm_1^4-\dm_2^4\right)-\left(\dm_1^2-\dm_3^2\right) \left(\dm_1^4-\dm_3^4\right)\right)
\right.
\nonumber \\ &\qquad\quad\, 
\left.
+
4 \cos 2 \th_{13R} 
\left(
4 \left[
\left(\dm_1^2-\dm_2^2\right) \left(\dm_1^4-\dm_2^4\right)
+\left(\dm_1^2-\dm_3^2\right) \left(\dm_1^4-\dm_3^4\right)
\right.\right.\right.
\nonumber \\ &\qquad\qquad\qquad\qquad\qquad
\left.\left.\left.
+\cos 2 \th_{23R} \left(
 \left(\dm_1^2-\dm_2^2\right) \left(\dm_1^4-\dm_2^4\right)
-\left(\dm_1^2-\dm_3^2\right) \left(\dm_1^4-\dm_3^4\right)
\right)
\right]
\right.\right.
\nonumber \\ &\qquad\qquad\qquad\qquad\,\,\,\,\,
\left.\left.
-(3 + 5 \cos 4 \th_{23R}) \left(\dm_2^2-\dm_3^2\right) \left(\dm_2^4-\dm_3^4\right) \right)
\right.
\nonumber \\ &\qquad\quad\, 
\left.
+
2 \cos 4 \th_{13R} 
\left(
2 \left[
 \cos 2 \th_{23R} \left(
  \left(\dm_1^2-\dm_3^2\right) \left(\dm_1^4-\dm_3^4\right)
 -\left(\dm_1^2-\dm_2^2\right) \left(\dm_1^4-\dm_2^4\right)
\right) 
\right.\right.\right.
\nonumber \\ &\qquad\qquad\qquad\qquad\qquad 
\left.\left.\left.
+ \left(\dm_1^2-\dm_2^2\right)\left(\dm_1^4-\dm_2^4\right)
+ \left(\dm_1^2-\dm_3^2\right)\left(\dm_1^4-\dm_3^4\right)
\right]
\right.\right.
\nonumber \\ &\qquad\qquad\qquad\qquad\,\,\,\,\,
\left.\left.
-\sin^2 2 \th_{23R} \left(\dm_2^2-\dm_3^2\right) \left(\dm_2^4-\dm_3^4\right) \right)
\right]\,,\\
%
%
\Delta^{(2)}_6 &=
\frac{1}{16}\,
\frac{\tm_1^2 - \tm_2^2}{\tm_1 \tm_2}\,
\sin 4 \th_{12R}\, \sin \th_{13R}\,\sin 2 \th_{23R}
\left(\dm_2^2-\dm_3^2\right)
\nonumber \\ &\qquad \times
\left[
(3-\cos 2 \th_{13R})\cos 2 \th_{23R} \left(\dm_2^4-\dm_3^4\right) 
\right.
\nonumber \\ &\qquad\quad\, 
\left.
-2\cos^2 \th_{13R} \left(
\left( \dm_1^4-\dm_2^4\right)
+\left(\dm_1^4-\dm_3^4\right)
-\left( \dm_1^2-\dm_2^2\right)\left(\dm_1^2-\dm_3^2\right)
 \right) 
\right]\,,\\
%
%
\Delta^{(2)}_7 &=
\frac{1}{16}\,
\frac{\tm_1^2 - \tm_2^2}{\tm_1 \tm_2}\,
\left(3+\cos 4 \th_{12R}\right)\,\sin^2 \th_{13R}\,\sin^2 2\th_{23R}
\left(\dm_2^2-\dm_3^2\right)\left(\dm_2^4-\dm_3^4\right)\,,\\
%
%
\Delta^{(2)}_8 &=
-\frac{1}{8}\,
\frac{\tm_1^2 - \tm_3^2}{\tm_1 \tm_3}\,
\cos \th_{13R}\, \sin 2\th_{13R}\,\sin 2\th_{23R}\,\sin 2\th_{12R}
\left(\dm_2^2-\dm_3^2\right)
\nonumber \\ &\qquad \times
\left[
 \left(\dm_1^4-\dm_2^4\right)
+\left(\dm_1^4-\dm_3^4\right)
-\left(\dm_1^2-\dm_2^2\right)\left(\dm_1^2-\dm_3^2\right)
+ \cos 2 \th_{23R}\left(\dm_2^4-\dm_3^4\right)
\right]\,,\\
%
%
\Delta^{(2)}_9 &=
\frac{1}{4}\,
\frac{\tm_1^2 - \tm_3^2}{\tm_1 \tm_3}\,
\cos^2 \th_{13R}\,\sin^2 \th_{12R}\,\sin^2 2\th_{23R}
\left(\dm_2^2-\dm_3^2\right)\left(\dm_2^4-\dm_3^4\right)\,.
\end{align*}

\begin{align*}
%
%
\Delta_{1}^{\prime\, (a)} &=
-\frac{1}{8}\,
\dm_1\,\dm_3\left(\dm_1^{2a}-\dm_3^{2a}\right)\,
\cos \th_{12L}\, \sin 2 \th_{13L}\,
\cos \th_{13R}\, \sin 2\th_{12R}\, \sin \th_{23R}
\nonumber \\ &\qquad \times
\left[
 \left(\hle-\hlm\right)
+\left(\hle-\hlt\right)
+ \cos 2 \th_{23L}\left(\hlm-\hlt\right)
\right] \left(\tm_1^2-\tm_2^2\right)\,,\\
%
%
\Delta_{2}^{\prime\, (a)} &=
\frac{1}{16}\,
\dm_2\,\dm_3\left(\dm_2^{2a}-\dm_3^{2a}\right)\,
\cos \th_{12L}\, \cos \th_{13L}\, \sin 2 \th_{23L}\, \sin 2 \th_{23R}
\nonumber \\ &\qquad \times
\left[
 2\cos^2 \th_{13R} \left(
\left(\tm_1^2-\tm_3^2\right)+\left(\tm_2^2-\tm_3^2\right)
\right)
\right.
\nonumber \\ &\qquad\qquad\,\, 
\left.
-\cos 2 \th_{12R}\left(3- \cos 2 \th_{13R}\right) \left(\tm_1^2-\tm_2^2\right)
\right] \left(\hlm-\hlt\right)\,,\\
%
%
\Delta_{3}^{\prime\, (a)} &=
-\frac{1}{8}\,
\dm_2\,\dm_3\left(\dm_2^{2a}-\dm_3^{2a}\right)\,
\sin \th_{12L}\, \sin 2 \th_{13L}\,
\sin 2\th_{12R}\, \sin \th_{13R}
\nonumber \\ &\qquad \times
\left[
 \left(\hle-\hlm\right)
+\left(\hle-\hlt\right)
+ \cos 2 \th_{23L}\left(\hlm-\hlt\right)
\right] \left(\tm_1^2-\tm_2^2\right)\,,\\
%
%
\Delta_{4}^{\prime\, (a)}&=
-\frac{1}{8}\,
\dm_1\,\dm_2\left(\dm_1^{2a}-\dm_2^{2a}\right)\,
\sin \th_{13L}\, \sin 2 \th_{23L}\,
\sin 2\th_{13R}\, \sin \th_{23R}
\nonumber \\ &\qquad \times
\left[
\left(\tm_1^2-\tm_3^2\right)+\left(\tm_2^2-\tm_3^2\right)
+ \cos 2 \th_{12R}\left(\tm_1^2-\tm_2^2\right)
\right] \left(\hlm-\hlt\right)\,,\\
%
%
\Delta_{5}^{\prime\, (a)} &=
\frac{1}{8}\,
\dm_1\,\dm_3\left(\dm_1^{2a}-\dm_3^{2a}\right)\,
\cos \th_{13L} \, \sin \th_{12L}\, \sin 2 \th_{23L}\,
\cos \th_{23R} \, \sin 2\th_{13R}
\nonumber \\ &\qquad \times
\left[
\left(\tm_1^2-\tm_3^2\right)+\left(\tm_2^2-\tm_3^2\right)
+ \cos 2 \th_{12R}\left(\tm_1^2-\tm_2^2\right)
\right] \left(\hlm-\hlt\right)\,,\\
%
%
\Delta_{6}^{\prime\, (a)} &=
-\frac{1}{32}\,
\dm_2\,\dm_3\left(\dm_2^{2a}-\dm_3^{2a}\right)\,
\sin \th_{12L}\, \sin 2 \th_{13L}\, \sin 2 \th_{23R}
\nonumber \\ &\qquad \times
\left[
 2\cos^2 \th_{13R} \left(
\left(\tm_1^2-\tm_3^2\right)+\left(\tm_2^2-\tm_3^2\right)
\right)
-\cos 2 \th_{12R}\left(3- \cos 2 \th_{13R}\right) \left(\tm_1^2-\tm_2^2\right)
\right] 
\nonumber \\ &\qquad \times
\left[
 \left(\hle-\hlm\right)
+\left(\hle-\hlt\right)
+ \cos 2 \th_{23L}\left(\hlm-\hlt\right)
\right] \left(\hlm-\hlt\right)\,,\\
%
%
\Delta_{7}^{\prime\, (a)} &=
-\frac{1}{4}\,
\dm_1\,\dm_2\left(\dm_1^{2a}-\dm_2^{2a}\right)\,
\cos 2\th_{12L}\, \sin \th_{13L}\, \sin 2\th_{23L}\,
\cos \th_{13R}\, \cos \th_{23R}\, \sin 2\th_{12R}
\nonumber \\ &\qquad \times
\left(\hlm-\hlt\right)
\left(\tm_1^2-\tm_2^2\right)\,, \\
%
%
\Delta_{8}^{\prime\, (a)} &=
\frac{1}{16}\,
\dm_1\,\dm_2\left(\dm_1^{2a}-\dm_2^{2a}\right)\,
\sin 2\th_{12L}\,
\cos \th_{13R}\,\cos \th_{23R}\,\sin 2\th_{12R}
\left(\tm_1^2 - \tm_2^2\right)
\nonumber \\ &\qquad \times
\left[
 2\cos^2 \th_{13L} \left(
\left(\hle-\hlm\right)+\left(\hle-\hlt\right)
\right)
-\cos 2 \th_{23L}\left(3- \cos 2 \th_{13L}\right) \left(\hlm-\hlt\right)
\right] \,, \\
%
%
\Delta_{9}^{\prime\, (a)} &=
\frac{1}{32}\,
\dm_1\,\dm_2\left(\dm_1^{2a}-\dm_2^{2a}\right)\,
\sin 2\th_{12L}\,
\sin 2\th_{13R}\,\sin \th_{23R}
\nonumber \\ &\qquad \times
\left[
 2\cos^2 \th_{13L} \left(
\left(\hle-\hlm\right)+\left(\hle-\hlt\right)
\right)
-\cos 2 \th_{23L}\left(3- \cos 2 \th_{13L}\right) \left(\hlm-\hlt\right)
\right]
\nonumber \\ &\qquad \times
\left[
\left(\tm_1^2-\tm_3^2\right)+\left(\tm_2^2-\tm_3^2\right)
+ \cos 2 \th_{12R}\left(\tm_1^2-\tm_2^2\right)
\right]
 \,, \\
%
%
\Delta_{10}^{\prime\, (a)} &=
\frac{1}{4}\,
\dm_2\,\dm_3\left(\dm_2^{2a}-\dm_3^{2a}\right)\,
\cos \th_{12L}\, \cos \th_{13L}\, \sin 2\th_{23L}\,
\cos 2\th_{23R}\, \sin 2\th_{12R}\, \sin \th_{13R}
\nonumber \\ &\qquad \times
\left(\hlm-\hlt\right)
\left(\tm_1^2-\tm_2^2\right)\,,\\
%
%
\Delta_{11}^{\prime\, (a)} &=
\frac{1}{4}\,
\dm_1\,\dm_3\left(\dm_1^{2a}-\dm_3^{2a}\right)\,
\cos \th_{13L}\, \sin \th_{12L}\, \sin 2\th_{23L}\,
\cos \th_{13R}\, \sin 2\th_{12R}\, \sin \th_{23R}
\nonumber \\ &\qquad \times
\left(\hlm-\hlt\right)
\left(\tm_1^2-\tm_2^2\right)\,,\\
%
%
\Delta_{12}^{\prime\, (a)} &=
\frac{1}{16}\,
\dm_1\,\dm_3\left(\dm_1^{2a}-\dm_3^{2a}\right)\,
\cos \th_{12L}\,\sin 2\th_{13L}\,
\cos \th_{23R}\,\sin 2\th_{13R}
\nonumber \\ &\qquad \times
\left[
 \left(\hle-\hlm\right)
+\left(\hle-\hlt\right)
+ \cos 2 \th_{23L}\left(\hlm-\hlt\right)
\right]
\nonumber \\ &\qquad \times
\left[
\left(\tm_1^2-\tm_3^2\right)+\left(\tm_2^2-\tm_3^2\right)
+ \cos 2 \th_{12R}\left(\tm_1^2-\tm_2^2\right)
\right] \,.
\end{align*}

}

\end{document}